\documentclass[journal]{IEEEtran}

\usepackage{epsfig}

\usepackage{color}
\usepackage{latexsym,amssymb,amsmath,setspace,array,multirow}
\usepackage{algorithmic}
\usepackage{algorithm}
\usepackage{graphicx}
\usepackage{booktabs}
\usepackage{cases}
\usepackage{color}
\usepackage{subfig}
\usepackage{caption}
\usepackage[colorlinks,linkcolor=red,anchorcolor=red,citecolor=red]{hyperref}
\definecolor{Red}{rgb}{1,0,0}
\usepackage{makecell}
\usepackage{mathrsfs}
\usepackage{amsbsy}
\usepackage{comment}
\usepackage{ulem}
\usepackage[figuresright]{rotating}

\begin{document}
	%
	\title{A Koopman-based Deep Convolutional Network for Modeling Latent Dynamics from Pixels}
	
	%
	\author{Yongqian~Xiao*, 
		~Zixin~Tang*,
		~Xin~Xu,
		~Lilin~Qian
		\thanks{Yongqian Xiao, Zixin Tang, Xin Xu are with the College of Intelligence Science and Technology, National University of Defense Technology, Changsha 410073, China. email: (xuxin\_mail@263.net)}
		\thanks{Lilin Qian is with the Unmanned System Technology Research Center, NIIDT, Beijing 010, China.}
	}
	
	\markboth{Journal of \LaTeX\ Class Files}
	{Shell \MakeLowercase{\textit{et al.}}: Deep Neural Networks with Koopman Operator for Data-driven Modeling of Vehicle Dynamics}
	
	%



	\IEEEtitleabstractindextext{%
		\begin{abstract}
			With the development of end-to-end control based on deep learning, it is important to study new system modeling techniques to realize dynamics modeling with high-dimensional inputs. In this paper, a novel Koopman-based deep convolutional network, called CKNet, is proposed to identify latent dynamics from raw pixels. CKNet learns an encoder and decoder to play the role of the Koopman eigenfunctions and modes, respectively. The Koopman eigenvalues can be approximated by eigenvalues of the learned state transition matrix. The deterministic convolutional Koopman network (DCKNet) and the variational convolutional Koopman network (VCKNet) are proposed to span some subspace for approximating the Koopman operator respectively.
			Because CKNet is trained under the constraints of the Koopman theory, the identified latent dynamics is in a linear form and has good interpretability. Besides, the state transition and control matrices are trained as trainable tensors so that the identified dynamics is also time-invariant. We also design an auxiliary weight term for reducing multi-step linearity and prediction losses. Experiments were conducted on two offline trained and four online trained nonlinear forced dynamical systems with continuous action spaces in Gym and Mujoco environment respectively, and the results show that identified dynamics are adequate for approximating the latent dynamics and generating clear images. Especially for offline trained cases, this work confirms CKNet from a novel perspective that we visualize the evolutionary processes of the latent states and the Koopman eigenfunctions with DCKNet and VCKNet separately to each task based on the same episode and results demonstrate that different approaches learn similar features in shapes.
		\end{abstract}
		
		}

	\maketitle

	\IEEEdisplaynontitleabstractindextext

	%
	\IEEEpeerreviewmaketitle
	
	\section{Introduction}   \label{doa:sec:introduction}
	Data-driven modeling and control have been an important research topic in recent years. The aim is to realize high-performance modeling and control for complex systems which are difficult for traditional methods based on analytic modeling and control design. Among the recent advances in data-driven modeling and control, Koopman-based model identification methods have attracted increasing attention. As a class of system modeling methods based on the theory of Koopman operator, Koopman-based modeling can establish linearized dynamic models for nonlinear systems and it can approximate nonlinear systems globally in the infinite invariant subspace \cite{Christof2017On}\cite{2017On}. 
	\par
	In general, previous Koopman-based modeling approaches can be divided into two main classes. One is dynamic mode decomposition (DMD)\cite{schmid2010dynamic} which uses singular value decomposition (SVD) to extract intrinsic features for approximating the Koopman eigenvalues and modes with the eigenvalues and their corresponding eigenvectors. Family of DMD includes Exact DMD \cite{tu2013dynamic}, Hankel-DMD \cite{arbabi2017ergodic}, HAVOK\cite{brunton2017chaos}, Windowed DMD\cite{hemati2014dynamic}, KDMD \cite{kevrekidis2016kernel}, and moving horizon Hankel-DMD (MH-HDMD) \cite{avila2020data}. The other is extended dynamic mode decomposition (EDMD) \cite{williams2015data} which transforms the modeling process into a supervised learning problem and solves it with least-squares methods. Until now, Koopman-based approaches have been applied to approximate system dynamics in many fields, such as  fluid dynamics \cite{mezic2013analysis}, power system \cite{susuki2016applied}\cite{netto2018robust}, molecular conformation analysis \cite{klus2018kernel}, robotic systems \cite{mamakoukas2019local}, etc. Prior works \cite{proctor2018generalizing}\cite{williams2016extending} extended DMD and EDMD to forced dynamics so that linear control theorems can be applied for control \cite{korda2018linear}.	\par
	To realize approximation in finite-dimension, one feasible way is to construct a linear operator in a higher-dimensional space to approximate the infinite Koopman operator where the higher-dimensional representation is created by lifting the original state space via designing dictionary functions. The choice of dictionary functions has a decisive effect on modeling performance. They can be constructed with kernel functions, e.g. radial basis functions (RBF), and Hermite polynomials. Unfortunately, designing dictionary functions demands strong experience and lacks of theoretical guidance. Besides, when the state dimension and the scale of dataset are extraordinarily large, it is intractable to load all data into memory and execute the SVD or pseudo-inverse operator. Under these circumstances, it is almost impossible to design suitable dictionary functions for systems whose states consist of high-dimensional pixels.
	
	In the past decades, deep learning algorithms or deep neural networks (DNNs) have shown promising capabilities in complex function approximation problems. Deep neural networks can be trained to take the place of dictionary functions instead of manually selecting different kernel functions with different hyper-parameters. From this perspective, some research works have been done on Koopman-based modeling approaches with deep neural networks, which have been applied in some fields such as fluid dynamics \cite{morton2018deep}, power grid \cite{ping2020deep}, vehicle dynamics \cite{xiao2020deep}, molecular kinetics \cite{mardt2018vampnets}, atomic scale dynamics \cite{xie2019graph} et.al. These works usually adopt an anto-encoder (AE) framework, and the encoder is used to approximate the Koopman eigenfunctions. Besides, some works approximated the system and control matrices according to the sequence of the latent states estimated by the encoder \cite{morton2019Deep}\cite{otto2019linearly} while some works treated them as trainable weights \cite{xiao2020deep}\cite{lusch2018deep}.
	
	Despite of the above advances, previous researches mainly focused on data-driven modeling of low-dimensional dynamic systems. However, under many circumstances, it is difficult to acquire real-time intrinsic vector valued states, but high-dimensional pixel-wise observations can be obtained via low-cost cameras. Furthermore, pixel-wise observations, such as images or lidar point cloud data, usually include lots of invalid information and noises. To deal with these situations, learning-based algorithms are usually applied, such as learning-based nonlinear MPC (LB-NMPC) \cite{ostafew2016learning}, and end-to-end learning control methods based on deep reinforcement learning algorithms (DRL) \cite{mnih2015human}\cite{haarnoja2018soft}. It is necessary to improve the efficiency of end-to-end control by designing latent dynamics model with raw pixels. Recently, encoding-based approaches were proposed to learn a neural network model or construct an encoder to improve the learning efficiency, such as MuZero \cite{schrittwieser2020mastering}, CURL \cite{laskin2020curl}, and PlaNet \cite{hafner2019planet}. Nevertheless, CURL only extracts features as the states for RL algorithms without predictive ability while MuZero and PlaNet learn a model that estimates the latent states via the extracted nonlinear feature vectors from deep neural networks in an end-to-end style.
	
	This paper proposes a novel Koopman-based deep convolutional network, called CKNet, to identify latent dynamics from raw pixels. CKNet learns an encoder and decoder to play the role of the Koopman eigenfunctions and modes, respectively. The Koopman eigenvalues can be approximated by the learned state transition matrix. Since CKNet is trained under the constraints of the Koopman theory, the identified dynamics is linear, controllable and physically-interpretable. An auxiliary weight term is designed for improving long-term prediction ability. Different from MuZero and PlaNet, the proposed CKNet regards the feature vector as the latent system's state (are also the basis functions for spanning the subspace of observations) and adopts the EDMD theory to approximate the Koopman operator of the latent system so that an interpretable linear dynamics can be established instead of a non-linear end-to-end model based on neural networks. In particular, although the latent state does not correspond to physical representation, it is a certain state for a certain linear system. In this way, the identified dynamics can predict for a long term in a linear form with a small computing cost.
	
	Currently, there are few works that focus on approximating the Koopman operator of dynamical forced systems with raw pixels as the state. In  \cite{haq2019universal}, a DMD-based deep learning framework was constructed for background/foreground extraction and video classification. Firstly, it focused on unforced systems. Secondly, it adopts a hierarchical manner that trains AE firstly then does the DMD procedure.
	DeepKoCo \cite{van2020deepkoco} learns a latent linear time-varying system from pixels. The encoder adopts a deterministic approach to output the Koopman eigenfunctions directly, and the system and control matrices are equal to the gradients of the Koopman eigenfunctions to the observation and action. Thus it does not assume the latent dynamics is linear in the control. Different from DeepKoCo, CKNet obtains a latent linear controllable time-invariant dynamics. Namely, the system and control matrices are fixed after training so that the identified dynamics is more suitable for controller design.
	
	In CKNet, the EDMD theory is adopted to design the neural network framework for approximating the Koopman operator. The encoder is realized with both DCKNet and VCKNet separately to output the latent states which is linear correlation with the Koopman eigenfunctions where the weight matrix consists of eigenvectors. The state transition and control matrices are dealt with trainable tensors. Namely, after training, the obtained latent linear dynamics is time-invariant, and controllability analysis can be done on the identified latent dynamics. The main contributions of this work are three-fold:
	\begin{itemize}
		\itemindent 2.8em
		\item[(1)] A convolutional neural network based on the Koopman operator is proposed and realized with the DCKNet and VCKNet for modeling latent dynamics from raw pixels.
		\item[(2)] An auxiliary weight term is proposed for multi-step linearity and prediction losses to improve the long-term prediction performance.
		\item[(3)] CKNet is applied to identify two systems in Gyms and four systems in Mujoco with continuous action space. For each task in Gym, we analyze the Koopman eigenfunctions based on the same episode with the DCKNet and VCKNet separately.
	\end{itemize}
	
	The rest of the paper is formed as follows: In Sec. \ref{sec:CKNet_design}, CKNet is designed for identifying unforced and forced dynamics. In Sec. \ref{sec:experiments} and Sec. \ref{sec:results}, two cases in Gym and four cases in Mujoco with continuous action space are identified to provide experimental validations for the proposed CKNet. Finally, conclusions and perspectives on the future are drawn in Sec. \ref{sec:conclusions}.
	
	\section{The Koopman-based CNN for Modeling Latent Dynamics with Raw Pixel}\label{sec:CKNet_design}
	For complex high-dimension systems, an accurate and efficient model is important for planning and controlling.
	In this section, the detailed process is presented on how to design CKNet for approximating the Koopman operator of discrete-time unforced and forced dynamical systems that take pixel-wise matrices as states. Also, the sampling method for VCKNet is given.
	\subsection{CKNet for Unforced Systems}
	Consider an unforced discrete-time system as follows:
	\begin{equation}\label{equ:unforced_dynamics}
	x_{k+1}=f\left( x_k \right)
	\end{equation}
	where $x \in \mathbb{R}^{c \times h \times w} \in \mathcal{M}$ denotes the state of dynamical system $f$ in original high-dimension space $\mathcal{M}$ and it consists of $c$ images with height $h$ and width $w$. 
	The Koopman operator is a linear operator in some infinite-dimensional space $\mathcal{H}$ spanned by the Koopman eigenfunctions, and the Koopman operator $\mathcal{K}$ can be defined by
	\begin{equation}
	(\mathcal{K}\varphi)(x_k) = \varphi(f(x_k))
	\end{equation}
	where $\varphi = [\varphi_1  \ \varphi_2  \ ...  \ \varphi_v]^{\top} \in \mathcal{H}$ are the Koopman eigenfunctions. In this manner, the unforced system $f$ can be described as a new dynamical system in $\mathcal{H}$:
	\begin{equation}\label{equ:classic_Koopman}
	\varphi \left( x_{k+p} \right) =(\mathcal{K}^p\varphi _i)(x_k)=\varLambda ^p\varphi \left( x_k \right)
	\end{equation}
	where $\mathcal{K}^p$ denotes $p$ times Koopman operator, $\Lambda$ is the diagonal matrix consists of the Koopman eigenvalues. Though $\mathcal{K}$ operator acts in infinite-dimensional space, it attracts attention because of its linearity. Through the thought of dimensionality reduction, methods based on DMD approximate the Koopman operator by approximating the Koopman eigenvalues and modes with SVD directly. EDMD needs to design dictionary functions manually, but it is difficult to choose suitable dictionary functions. Generally speaking, since DMD does not approximate the Koopman eigenfunctions, EDMD can obtain better results if appropriate dictionary functions are chosen. But for systems with pixels as inputs, classic DMD and EDMD are unable to cope with such high-dimensional and large-scale problems. Furthermore, nonlinear systems are expressed in a linear manner, and new states are embodied with vector-valued basis functions no longer with high-dimensional pixels, but it is intractable because of its infinity of dimensions.	\par
	In this work, a deep learning framework based on EDMD is proposed to approximate the Koopman operator to obtain better performance.	A CNN encoder is utilized to automatically learn finite-dimensional dictionary functions which have two purposes. The one is to extract features from the inputted pixels, and the other is to project these features to some subspace of $\mathcal{H}$ so that a finite dimensional approximation of the Koopman operator can be generated.	\par

\begin{figure*}[htbp]
		\begin{center}
			\includegraphics[scale=0.4]{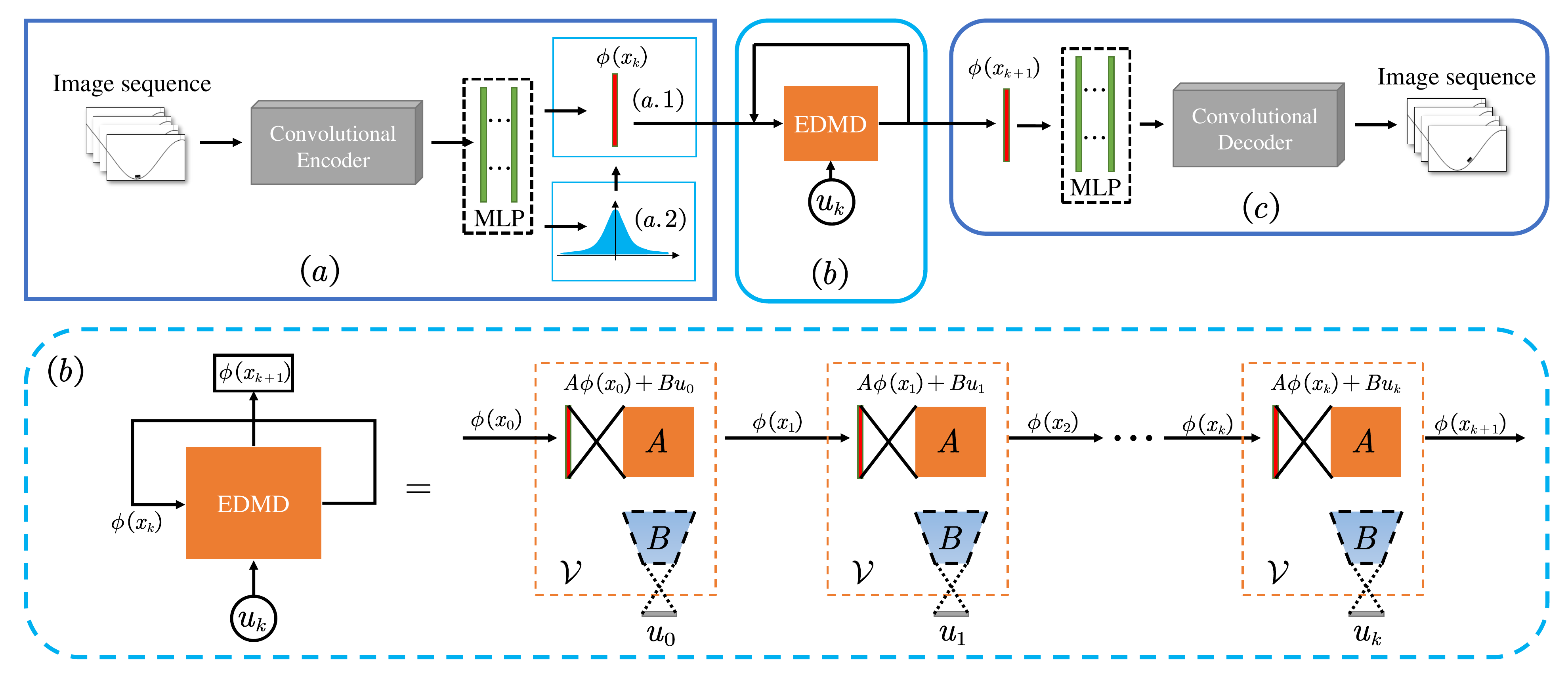}
			\caption{The framework of CKNet. (a) The encoder of CKNet for expanding some finite space of $\mathcal{V}$ as an invariant subspace of $\mathcal{H}$. It outputs basis functions for playing the role of the Koopman eigenfunctions. The encoder is constructed in two ways, the DCKNet and VCKNet respectively. The DCKNet shown in (a.1), outputs basis functions directly after the MLP. (a.2) shows the VCKNet via sampling from the learned Gaussian distribution. (b) We adopt a recursive way to realize multi-step training. A high-dimensional nonlinear systems can be described as a low-dimensional linear dynamics $\phi \left( x_{k+1} \right) =A\phi \left( x_k \right) +Bu_k$ in $\mathcal{V}$, where the state transition matrix $A$ and control matrix $B$ are obtained as trainable tensors. CKNet is also applicable for unforced systems while the input $u_k$ equals 0 constantly. (c) The decoder has the reverse structure with the encoder and it plays the role of the Koopman modes for mapping the latent states from subspace $\mathcal{V}$ back to the original observation space.}
			\label{fig:DCKNet_constructure}
		\end{center}
	\end{figure*}

	As shown in Fig. \ref{fig:DCKNet_constructure}, CKNet expands  a low-dimensional subspace $\mathcal{V}$ via the encoder $\phi$ to extract intrinsic dynamical features as the latent states to play the role of the Koopman eigenfunctions. Meanwhile, a nonlinear CNN decoder is designed to play the role of the linear Koopman modes to transform the latent states from $\mathcal{V}$ back to pixels in $\mathcal{H}$.
	Therefore, the unforced system $f$ can be approximated via CKNet:
	\begin{equation}\label{equ:DCKNet_unfoced_dynamics}
	\begin{cases}
	\phi \left( x_{k+p} \right) \doteq \phi _{\mathscr{K}}(x_{k+p})=\mathscr{K}^p\phi \left( x_k \right) =A^p\phi \left( x_k \right)\\
	x_k= \tilde{\phi}\left( \phi(x_k) \right)\\
	\end{cases}
	\end{equation}
	where $\mathscr{K}$ is a finite approximating operator of $\mathcal{K}$ in $\mathcal{V}$ and represented by the square matrix $A \in \mathcal{R}^{v \times v}$ while $A^p$ denotes $p$ times $\mathscr{K}$ operator recurrently, $\phi(x) \in \mathbb{R}^{v}$ and $\phi_{\mathscr{K}}(x) \in \mathbb{R}^{v}$ denote the latent state acquired via the encoder and $\mathscr{K}$ operator, respectively, $\tilde{\phi}(\phi(x)) \in \mathbb{R}^{c' \times h \times w}$ denotes the output of the decoder, where $c'$ is a hyper-parameter which equals $c$ or 1. Note that basis functions are the latent states in this work, and (\ref{equ:DCKNet_unfoced_dynamics}) is a new dynamics even though it is used to approximate nonlinear systems. \par
	With the DCKNet, the encoder outputs basis functions $\phi(x)$ directly. When the encoder adopts the VCKNet, the encoder outputs the mean and logarithm of variance to construct a Gaussian distribution so that basis functions can be acquired by sampling from this distribution. To realize back-propagation, reparameterized technique is applied to sample basis function in the training process:
	\begin{equation}\label{equ:bf_sample}
	\phi \left( x \right) =\mu _{\phi}\left( x \right) +\exp \left( \ln \left( \sigma _{\phi}\left( x \right) \right) \odot \xi \right)
	\end{equation}
	where $\mu_{\phi}$ and $\sigma _{\phi}$ are the mean and variance of the learned Gaussian distribution. The $\mu$ and logarithm of variance $\ln(\sigma_{\phi})$ are given by the variational encoder directly. Where $\odot$ denotes dot product, $\xi \sim \mathcal{N}\left( 0, I \right)$ is a noise vector from a standard normal distribution.
	
	In this work, we adopt a fixed sampling interval to sample datasets and learned a discrete system to describe the original continuous time system. After training, the Koopman eigenvalues are approximated by the eigenvalues of $A$. At the same time, the Koopman eigenvalues of the original continuous time system can be approximated by $\lambda _i=\frac{\ln \left( \mu _i \right)}{\varDelta t}$, where $\varDelta t$ is the sampling interval.
	Basis functions are induced by the encoder and linear correlation with the Koopman eigenfunctions. Especially, weight for calculating the $i-th$ Koopman eigenfunction can be acquired with the learned system matrix: 
	\begin{equation}\label{equ:functions_encoder}
	\varphi _i=\boldsymbol{w}_{i}^{*}\phi \left( x \right)
	\end{equation}
	where $\boldsymbol{w}_{i}$ is the left eigenvector corresponding to the $i$-th eigenvalue.
	The Koopman eigenvalues and eigenfunctions are intrinsic features of systems, but the Koopman modes are determined by the Koopman eigenfunctions to realize the full state observation. Namely, the purpose of the Koopman modes are to remap the Koopman eigenfunctions from the subspace $\mathcal{V}$ back to the original state space $\mathcal{M}$ though the identifying accuracy depends on the training performance of the encoder and system matrices. For vector valued cases, the Koopman modes can be obtained via constructing least square problem based on the learned basis functions:
	\begin{equation}\label{equ:modes_for_lowD_cases}
	\begin{cases}
	x=\mathcal{B}\phi \left( x \right) \\
	x=\zeta \varphi
	\end{cases}	
	\end{equation}
	where $\mathcal{B}=\left[ \boldsymbol{b}_1\,\,\cdots \,\,\boldsymbol{b}_v \right] \in \mathcal{R}^{N \times v}$ is the weight matrix to remap the latent state to the original state, and $N$ is the dimension of the original vector valued dynamical system. For notational convenience, we present $W =\left[ \boldsymbol{w}_1\,\,\cdots \,\,\boldsymbol{w}_v \right] $ and $\varXi$ as the matrices consisting of left and right eigenvectors respectively. Therefore, there are $\varphi = W^* \phi$ and $\varXi^{-1}= W^*$. Combining (\ref{equ:functions_encoder}) and (\ref{equ:modes_for_lowD_cases}), the weight matrix $\mathcal{B}$ and Koopman modes can be calculated:
	\begin{equation}
	\begin{cases}\label{equ:B_modes}
	\mathcal{B}=X\phi \left( X \right) ^\top \left( \phi \left( X \right) \phi \left( X \right) ^\top \right) ^{-1} \\
	\zeta =\mathcal{B}\varXi
	\end{cases}
	\end{equation}
	where $X=[x_1\,\, \cdots \,\,x_M]\in \mathcal{R}^{N\times M}$ is the training dataset that consists of vector valued states. However, for pixel-wise systems, $\mathcal{B}$ can not be obtained by constructing the least square problem and the Koopman modes can not be represented in a linear style in (\ref{equ:B_modes}) anymore. Besides, we can not establish the least square problem between the Koopman eigenfunctions and original pixels to obtain the Koopman modes since we cannot express the Koopman modes in terms of a matrix to realize full observation by remapping vector-valued eigenfunctions to high-dimensional pixels.  In this situation, the CNN decoder takes the place of the Koopman modes to remap basis functions $\phi(x)$ to original pixels $\tilde{\phi}(\phi(x))$. Besides, the Koopman modes are not necessary for the predicting process. Further, though the prediction process can be executed with the approximated Koopman eigenvalues and eigenfunctions, the method in (\ref{equ:DCKNet_unfoced_dynamics}) is more effective because the step of eigenvalue decomposition on $A$ can be removed.
	
	\subsection{CKNet for Forced Dynamics}
	Compared to unforced dynamics, forced dynamics has an extra responding term to the control, but it is consistent in terms of framework designing. Consider a discrete-time forced dynamics:
	\begin{equation}\label{equ:forced_dynamics}
	x_{k+1}=f\left( x_k, u_k \right)
	\end{equation}
	where $x_k \in \mathbb{R}^{c \times h \times w}$ and $u_k \in \mathbb{R}^n$ are the state and control of the system $f$.
	There are several methods to extend the Koopman operator for forced systems which take value-wise vector as states\cite{proctor2018generalizing, williams2016extending, korda2018linear}. In this work, we design CKNet for forced dynamics based on the method in \cite{korda2018linear}. The Koopman operator of (\ref{equ:forced_dynamics}) can be described as follows:
	\begin{equation}\label{equ:koopman_forced_dynamics}
	\left( \mathcal{K}\varphi \right) \mathcal{X}_k=\varphi \left( f\left( \mathcal{X}_k \right) \right)
	\end{equation}
	where $\mathcal{X}$ is the extended state of the dynamics:
	\begin{equation}\nonumber
	\mathcal{X}_k=\left[ \begin{array}{c}
	x_k\\
	u_k\\
	\end{array} \right]
	\end{equation}
	Similarly, CKNet is also applicable for approximating the Koopman operator in (\ref{equ:koopman_forced_dynamics}):
	\begin{equation}\label{equ:DCKNet_forced_dynamics}
	\begin{cases}
	\phi \left( x_{k+p} \right) \doteq \phi _{\mathscr{K}}\left( x_{k+p} \right) =\mathcal{G}^p\varPsi \left( x_k \right)\\
	x_k=\tilde{\phi}\left( x_k \right)\\
	\end{cases}
	\end{equation}
	where $\mathcal{G}=\left[ A\,\,B \right] $ is an approximating operator for forced dynamics. $\varPsi \left( x_k \right) =\left[ \phi \left( x_k \right) \,\,u_k \right]^ \top$ is the extended state in $\mathcal{V}$. In EDMD, the system matrix $A$ and control matrix $B$ are solved by constructing a least square problem that builds the dataset with snapshot pairs. However, it is not feasible for large-scale dynamical systems. In this work, $A$ and $B$ are treated as trainable tensors and trained in a mini-batch manner.
	Particularly, we execute the controllability analysis of the identified dynamics in subspace $\mathcal{V}$ in the training process. The approximated discrete-time linear dynamics is controllable if $\mathcal{R}$ equals the dimension of the latent state, $\upsilon$:
	\begin{equation}\label{equ:controllable_analysis}
	\begin{cases}
	S=\left[ B\,\,AB\,\,A^2B\,\,... \ A^{\upsilon-1}B \right]\\
	\mathcal{R}=Rank\left( S \right)\\
	\end{cases}
	\end{equation}
	\subsection{Loss functions for CKNet}
	Different from previous works, CKNet is a general framework based on the EDMD theory and can be viewed as an extension of EDMD for forced dynamics modeling based on deep learning. Meanwhile, CKNet extends the scope of application mainly in three aspects: Firstly, CKNet adopts multi-step loss functions with the auxiliary term to improve the approximating performance. Secondly, CKNet is applicable for dynamical systems which task pixels as inputs. Thirdly, CKNet utilizes a mini-batch training manner with the DCKNet and VCKNet. \par
	To strength the linear accuracy of the identified model in $\mathcal{V}$, linearity loss is considered to restrain the encoder. Multi-step linearity loss technique is applied so that the approximated dynamics can be identified better in a global perspective leading to more predictive steps without divergence.
	\begin{equation}\label{equ:linear_loss}
	\begin{aligned}
	L_{linear}&=\frac{1}{p_l} \sum_{i=1}^{p_l}{\iota _i \left\| \phi \left( x_{k+i}, \theta_e \right) -\phi _{\mathscr{K}}\left( x_{k+i} \right) \right\| _{F}^{2}}\\
	&=\frac{1}{p_l}  \sum_{i=1}^{p_l}{\iota _i \left\| \phi \left( x_{k+i}, \theta_e \right) -\mathcal{G}^{i}\varPsi \left( x_k \right) \right\| _{F}^{2}}
	\end{aligned}
	\end{equation}
	where the encoder $\phi$ is parameterized with trainable weights $\theta_e$, $\iota_i$ is the auxiliary weight of the $i$-th step linear prediction, and it is defined by:
	\begin{equation}\label{equ:auxiliary_weight}
		\iota _i=1+\tanh \left( \tau _li \right)
	\end{equation}
	where $\tau_{l}$ is a hyper-parameter that decides the importance of a long-term linear evolution. Adopting the method of (\ref{equ:auxiliary_weight}), $\tau_{l}$ is limited in the range of $[1,\ 2]$ so that gradient explosion will not happen.
	$\mathcal{G}^i$ denotes $i$ times linear recursion from a state $\phi(x_k,\theta_e)$ with a sequence of control $u_k, ..., u_{k+i}$ in $\mathcal{V}$, and it can be calculated as follows:
	\begin{equation}\label{equ:multi_step_pred}
	\begin{aligned}
	\mathcal{G}^i\varPsi \left( x_0 \right) &=\phi _{\mathscr{K}}\left( x_{k+i} \right) \\
	&=A\phi _{\mathscr{K}}\left( x_{k+i-1} \right) +Bu_{k+i-1} \\
	&=\cdots \\
	&=A^i\phi \left( x_k,\theta _e \right) +\sum_{j=1}^i{A^{j-1}Bu_{k+i-j}}
	\end{aligned}
	\end{equation}
	\par
	When the encoder, $A$, and $B$ are trained only with the constraint (\ref{equ:linear_loss}), we can also acquire a linear approximated dynamics in $\mathcal{V}$ if we don't demand to obtain corresponding pixels. However, during the training process, this training method will make the encoder, $A$, and $B$ converge to zeros gradually which results in an invalid model. To avoid this problem and make sure features are extracted validly, a reconstruction loss function is included. This loss restrains the intrinsic features extracted by the encoder to contain all the information so that the decoder could retrieve original pixels.
	\begin{equation}\label{equ:reconstruction_loss}
	L_{recon}=\frac{1}{p} \sum_{i=1}^p{\left\| x_{k+i}-\tilde{\phi}\left( \phi \left( x_{k+i},\theta _e \right) ,\theta _d \right) \right\| _{F}^{2}}
	\end{equation}
	where $\tilde{\phi}$ denotes the decoder which is parameterized with $\theta_d$. \par
	Since we need to generate the corresponding images after multi-step prediction in $\mathcal{V}$, the weighted multi-step prediction loss function is designed to further restrain the encoder and decoder.
	\begin{equation}
	\label{equ:prediction_loss}
	L_{pred}=\frac{1}{p_p} \sum_{i=1}^{p_p}{\varrho _{i}\left\| x_{k+i}-\tilde{\phi}\left( \mathcal{G}^i\varPsi \left( x_k \right) ,\theta _d \right) \right\| _{F}^{2}}
	\end{equation}
	where $\varrho _i=1+\tanh \left( \tau _pi \right)$ is the auxiliary weight for the $i$-th step linear prediction. Similar to (\ref{equ:auxiliary_weight}), the hyper-parameter $\tau_p$ denotes the importance of a long-term reconstruction of linear evolution. Proper auxiliary weights for multi-step linearity and prediction loss could improve the quality of the learned dynamics. As a fact, the multi-step prediction loss $L_{pred}$ includes the multi-step linearity loss $L_{linear}$ and reconstruction loss $L_{recon}$ somehow. Therefore, only a small weight is needed for $L_{linear}$ in the training process. \par

	In addition, $l_2$ loss is added to avoid over-fitting.
	\begin{equation}\label{equ:l2_loss}
	l_2=\varTheta ^2
	\end{equation}
	where $\varTheta$ denotes the weights of the encoder, decoder, $A$, and $B$.
	Finally, CKNet can be trained under the loss function as follows:
	\begin{equation}\label{equ:total_loss}
	L=\alpha _1L_{linear}+\alpha _2L_{recon}+\alpha _3L_{pred}+\alpha _4l_2
	\end{equation}
	where $\alpha_1$, $\alpha_2$, $\alpha_3$, $\alpha_4$ are the weights for each loss function. CKNet can be trained through minimizing the weighted loss, $L$, and details are outlined in Algorithm \ref{alg:CKNet_alg}.
	
	\begin{algorithm}[h]
		\footnotesize
		\caption{The CKNet Algorithm}
		\label{alg:CKNet_alg}
		\begin{algorithmic}[1]
			\REQUIRE $p$, $p_l$, $p_p$, $\tau_p$ $\tau_l$, $c$, $c'$, $\zeta$, $\beta$,  $b_s$, $Epoch = 0$, $Epoch_{max}$, $\alpha_i$, $i=1,\cdots,4$.\\
			Initialize ${\theta}_e$, ${\theta}_d$, $A$, $B$, $ms=max(p, p_l, p_p)$.
			\ENSURE trained ${\theta}_e$, ${\theta}_d$, $A$, $B$;\\
			\WHILE {$Epoch <= Epoch_{max}$}
			\STATE Draw sequence data: $x_{0:ms}$, $u_{0:ms-1}$;
			\FOR{$i=0$ \TO $ms$ }
			\IF {Adopt dterministic approach}
			\STATE The encoder outputs $\phi(x_{i},\theta_e)$;
			\ELSE
			\STATE Sample the latent states $\phi(x_{i},\theta_e)$ with (\ref{equ:bf_sample});
			\ENDIF
			\STATE Reconstruct states $\hat{x}=\tilde{\phi}(\phi(x_{i},\theta_e),\theta_d)$;
			\ENDFOR
			\FOR{$i=1$ \TO $ms$}
			\STATE Compute $\mathcal{G}^{i}\varPsi(x_{0})$ and $\tilde{\phi}(\mathcal{G}^{i}\varPsi(x_{0}))$;
			\ENDFOR
			\STATE Obtain the weighted loss $L$ in (\ref{equ:total_loss}) with (\ref{equ:linear_loss}), (\ref{equ:auxiliary_weight}), (\ref{equ:reconstruction_loss}), (\ref{equ:prediction_loss}), and (\ref{equ:l2_loss});
			\STATE Update model parameters $\varTheta \gets \varTheta -\beta \nabla _{\varTheta}L\left( \varTheta \right)$
			\STATE $Epoch = Epoch + 1$
			\ENDWHILE
		\end{algorithmic}
	\end{algorithm}

	\section{Experimental Design}\label{sec:experiments}
	In this section, simulations on cases with continuous action spaces are conducted in offline and online training manner. For offline training cases, the MountainCar and CartPole tasks in Gym are adopted while cheetah-run, cartpole-balance, ball\_in\_cup-catch, and walker-walk are selected in Mujoco.
	\subsection{Cases in Gym}
\begin{figure}[htbp]
		\begin{center}
			\includegraphics[scale=0.45]{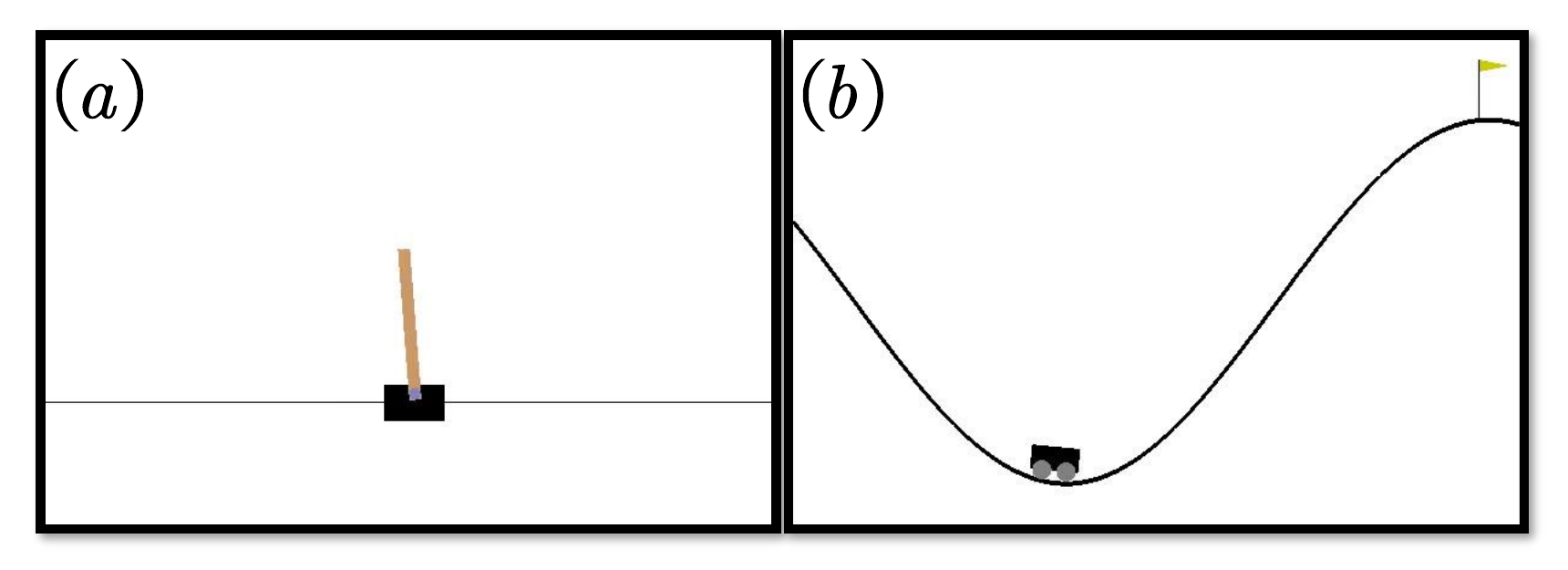}
			\caption{The selected two forced dynamics in Gym environment with continuous action space for validating CKNet. (a) The `CartPole' task; (b) The `MountainCarContinuous-v0' task.}
			\label{fig:tasks}
		\end{center}
	\end{figure}

	\textbf{Data collection: }
	As shown in Fig. \ref{fig:tasks}, `CartPole' and `MountainCarContinuous-v0' (MountainCar for brevity) are two classic tasks in Gym library \cite{brockman2016openai} for validating reinforcement learning (RL) algorithms. The analytic dynamics of these two systems are given in Sec. \ref{sec:analytic_dynamics}.
	
	In order to obtain comprehensive data in the state space, trained RL algorithms plus a term of noise were utilized as the controller for data collection.
	In the collection process, recorded episodes include the sequence of images $x_{k:T}$ and the executed action $u_{k:T-1}$. For CartPole, 250 episodes were collected, 25 for testing, 25 for validation, and the rest 200 for training. The episode length, $T$, is in the range of $[200,\ 300]$. For MountainCar, 240 episodes were collected, 20 for testing, 20 for validation, and the rest 200 for training. The steps of each episode are in the range of $[300,\ 400]$. \par
	In the preprocessing, images were firstly converted to grayscale. Then, these images were enhanced by modifying the grayscale to $1.0$ when a pixel's value is bigger than 0.8. Lastly, we crop and resize images to an appropriate size so that we can decrease the calculation cost but still keep enough key information.
	A single image includes position and angle features but it can not represent information of velocities, such as the velocity of the car and angular velocity of the pole in the CartPole task. Therefore, $c$ adjacent images are concatenated to a multi-channel tensor as the state. As shown in the first row in Fig. \ref{fig:images_result}, the size of state tensors is $c \times 460 \times 170$ for CartPole environment, $c \times 90 \times 90$ for MountainCar.
	
	\textbf{Training: }
	Neural network structures and activation functions are shown in Table. \ref{tab:structure}. The CartPole and MountainCar tasks have similar structures and activation functions. For the activation function of the encoder's last layer, we tried two kinds of activation styles, the `Tanh' function and without an activation function, and experiment results show that these two kinds of activation functions are both valid. Decoders have completely reversed neural network structures with corresponding to encoders. Activation functions of decoders are `ReLU' functions except the activation function of the last convolutional layer is `Sigmoid' function.\par
	\begin{table}[!t]
		\footnotesize
		\caption{Neural network structure of these two examples.}
		\centering
		\label{tab:structure}
		\begin{tabular}{cccc}
			\toprule
			\multicolumn{2}{c}{CartPole} & \multicolumn{2}{c}{MountainCar}\\
			\multicolumn{1}{c}{Structure} & \multicolumn{1}{c}{ACT}	& \multicolumn{1}{c}{Structure} & \multicolumn{1}{c}{ACT}\\
			\midrule
			$170 \times 460 \times 3$ & Input 	& $90 \times 90 \times 3$  & Input \\
			$84 \times 229 \times 8$  &  ReLU		& $44 \times 44 \times 16$ & ReLU \\
			$41 \times 113 \times 16$ &  ReLU		& $21 \times 21 \times 32$ & ReLU\\
			$19 \times 55 \times 32$  &  ReLU		& $9 \times 9 \times 64$ & ReLU\\
			$8 \times 26 \times 64$   & ReLU 		& $6 \times 6 \times 128$  & ReLU\\
			$3 \times 12 \times 128$  & ReLU		& - & - \\
			4860 & ReLU 							& 4860 & ReLU \\
			1525 & ReLU 							& 1525 & ReLU \\
			32 & -/Tanh							& 32   & -/Tanh \\
			\bottomrule
		\end{tabular}
	\end{table}
	Hyper-parameters are given in Table. \ref{tab:common_hp}, where $\beta$, $b_s$ are the learning rate and batch size respectively, $m$ and $n$ are intrinsic dimensions of tasks, $D-{\star}$, $V-{\star}$ denote the auxiliary weights for DCKNet and VCKNet respectively. For the MountainCar task, coefficients $\tau_l$ of the auxiliary weight for both DCKNet and VCKNet are equal to $0.03$. For the CartPole task, we do not introduce the auxiliary weight for the DCKNet while we adopt $\tau_p = 0.01$ for the VCKNet. For both CartPole and MountainCar, decent performances still can be obtained via setting coefficients of auxiliary weights to zero if we don't want to spend time tuning hyper-parameters. $c$ and $c'$ denote the number of images for the input of encoders and output of decoders. In the training, $c'$ could equals $c$ or 1 indicate different constraints from the output to the decoder. The value of $c'$ does not have obvious influences on identified results after testing.
	
	In the training process, CKNet does not need to design network structures and tune hyperparameters deliberately for different tasks. Additionally, batch-normalization technique is utilized after CNN layers. CKNet is trained with Pytorch-Lightning 1.0.7, which is a framework based on Pytorch and it is convenient for synchronizing parameters of batch-normalization with multi-GPU.\par
	\begin{table}[!t]
		\footnotesize
		\caption{Hyper-parameters of tasks in Gym.}
		\label{tab:common_hp}
		\centering
		\begin{tabular}{cccc}
			\toprule
			\multicolumn{1}{c}{Hyper-params} & \multicolumn{1}{c}{CartPole} & \multicolumn{1}{c}{MountainCar}\\	
			\hline
			$\alpha_1$ & 0.3 		& 0.3 \\
			$\alpha_2$ & 1.0 		 & 1.0\\
			$\alpha_3$ & 1.0 		 & 1.0\\
			$\alpha_4$ & $5\times 10^ {-7}$ & $5\times 10^ {-7}$\\
			$\upsilon$        & $32$ & $32$\\
			$m$	&$4$&	$2$ \\
			$n$&	$1$&	$1$	\\
			$p_l/p_p/p$ & $25$ & $25$\\
			$c$   & $3$  & $3$\\
			$c'$  & $3$  & $3$\\
			$\beta$  & $0.0004$ & $0.0001$\\
			$b_s$  & 16  & 32	\\
			$D-\tau_p$ & $0$ & $0$\\
			$D-\tau_l$ & $0$ & $0.03$\\
			$V-\tau_p$ & $0.01$ & $0$\\
			$V-\tau_l$ & $0$ & $0.03$\\
			\bottomrule
		\end{tabular}
	\end{table}

	\subsection{Cases in Mujoco}
\begin{figure}[htbp]
		\begin{center}
			\includegraphics[scale=0.24]{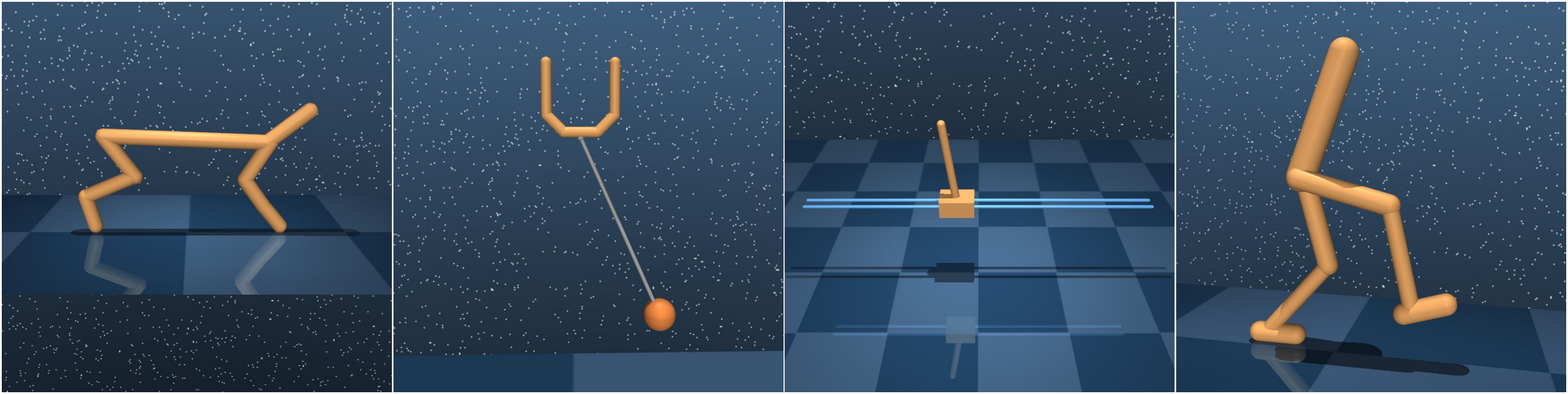}
			\caption{Adopted Mujoco cases. They are cheetah-run, ball\_in\_cup-catch, cartpole-balance, and walker-walk separately. And images are resized into $84\times 84$ as inputs.}
			\label{fig:mujoco_envs}
		\end{center}
	\end{figure}

	\textbf{Training: }
	Mujoco is a physics engine for control.	As shown in Fig. \ref{fig:mujoco_envs}, we adopt four cases in Mujoco to validate CKNet.
	\begin{table}[!t]
		\footnotesize
		\caption{Hyper-parameters of tasks in Mujoco.}
		\label{tab:common_hp_mujoco}
		\centering
		\begin{tabular}{cccccc}
			\toprule
			\multicolumn{1}{c}{Hyper-params} & \multicolumn{1}{c}{Ch-R} & \multicolumn{1}{c}{BIC-C}& \multicolumn{1}{c}{Cp-B}& \multicolumn{1}{c}{Wa-W}\\	
			\hline
			$\upsilon$        & $64$ & $128$ & $128$ & $128$\\
			$m$ & 18 & 8 & 4 & 18 \\
			$n$ & 6 & 2 & 1 & 6 \\
			$p_l/p_p/p$ & $20$ & $20$& $20$&$20$\\
			$c$   & $3$  & $3$& $3$ & $3$\\
			$c'$  & $3$  & $3$& $3$ & $3$\\
			$A_r$ & $4$	 & $4$	& $8$ & $2$	\\
			$b_s$  & 64  & 64	& 64 & 64 \\
			$\tau_{\star}$ & 0 & 0 & 0 & 0 \\
			\bottomrule
		\end{tabular}
	\end{table}

	The main common hyper-parameters are detailed in Table. \ref{tab:common_hp_mujoco}. For all four tasks, learning rates $\beta$ of actors, critics, encoders, and decoders are equal to $10^{-3}$ while learning rates of temperature and Koopmans are $10^{-4}$ and $5 \times 10^{-4}$ respectively. $A_r$ indicates numbers of repeating actions, i.e. executing the same action within $A_r$ steps. For Mujoco cases, we do not adopt auxiliary weights, that is, auxiliary weights $\tau_{\star} = 0, \star \in \{p, l\}$.
	
	The encoder is updated with the critic and Koopman operator while the decoder is only updated with the Koopman operator. The online training version of CKNet is realized based on the implement of SAC\footnote{\href{http://github.com/rail-berkeley/softlearning/}{http://github.com/rail-berkeley/softlearning/}} except that we utilize a replay buffer for sequences instead of transitions. All the neural network structures are the same with SAC except $A$ and $B$. Besides, because variational structures negatively influence SAC, and we did not train SAC successfully while we adopt VCKNet, we only utilize DCKNet to validate linear evolution prediction.

	\section{Results}\label{sec:results}
	\subsection{Cases in Gym}
\begin{figure}[htbp]
		\begin{center}
			\includegraphics[scale=1.]{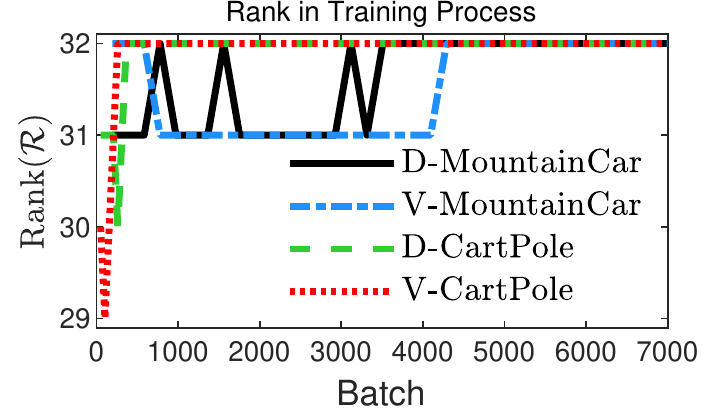}
			\caption{The rank in the training process of the matrix $S$ in (\ref{equ:controllable_analysis}). $D-\star$ denotes the task $\star$ realized by the DCKNet, and where $\star \in \{$ MountainCar, CartPole$\}$. Similarly, V-$\star$ denotes the task realized by the VCKNet.}
			\label{fig:rank}
		\end{center}
	\end{figure}

	\textbf{Controllability analysis: }
	During the training process, we regularly check the controllability of the identified linear dynamics by recording the rank of $S$ in (\ref{equ:controllable_analysis}). The change curves of the rank $\mathcal{R}$ are shown in Fig. \ref{fig:rank}, for the CartPole task, the rank $\mathcal{R}$ reaches $\upsilon$ quickly. For the MountainCar task, $S$ becomes full rank after around 4.5K training steps. Namely, in the training process, the identified dynamics of these two tasks for both the DCKNet and VCKNet are controllable. Note that we only show cases without auxiliary weights for the sake of charting brevity. For other trained cases with different auxiliary coefficients in $\{0.01,\ 0.03\}$, the obtained linear dynamics are also controllable. In addition, $ \tau_p $ and $ \tau_ {l} $ in all cases will not be assigned at the same time.	\par
\begin{figure*}[htbp]
	\centering
	\subfloat[Prediction results visualization of the CartPole task]{
		\label{subfig:cp_result}
		\includegraphics[width=\textwidth]{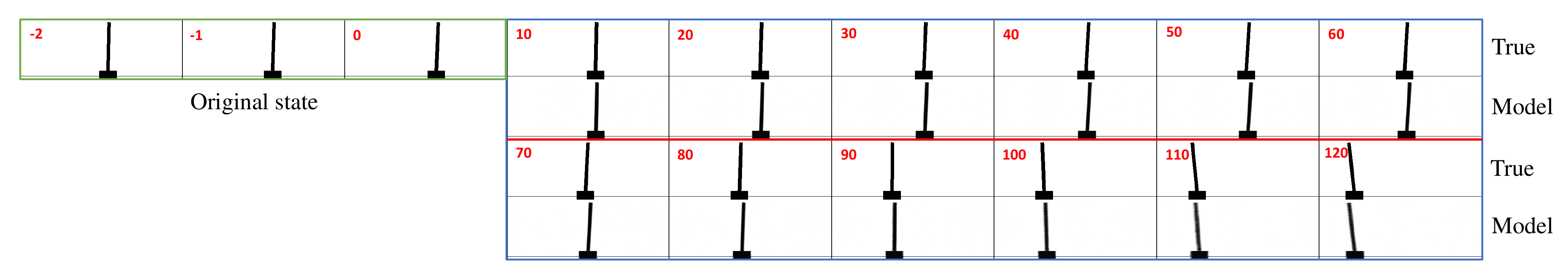}}\\
	\vspace{-0.1cm}
	\subfloat[Prediction results visualization of the MountainCar task]{
		\label{subfig:mc_result}
		\includegraphics[width=\textwidth]{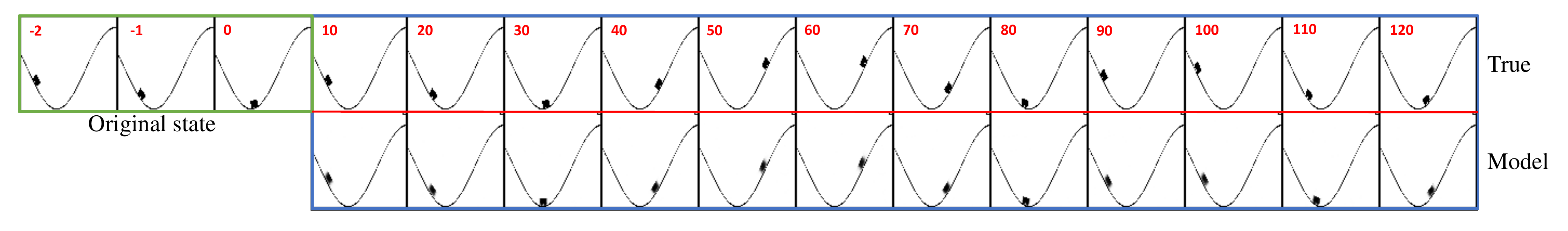}}\\
	\centering
	\caption{The `Model' rows are predicted by exerting a time sequence of actions on the original state $x_0$ which consists of three adjacent images. Firstly, obtain the original latent state $\phi(x_0)$, then the latent states of prediction horizon $\phi_{\mathscr{K}}(x_{1:120})$ can be obtained via linear evolution in subspace $\mathcal{V}$ with the sequence of action $u_{0:119}$. Finally, the predicted sequence of images $\hat{x}_{1:120}$ can be generated via the decoder with the latent states.}
	\vspace{-0.2cm}
	\label{fig:images_result}
\end{figure*}

	\textbf{Identification accuracy analysis: }
	As shown in Fig. \ref{fig:images_result}, $120$ steps predictions have been done for demonstrating the generated sequence of images via the proposed CKNet\footnote{\href{https://youtu.be/ZysYNBI_Y7w}{https://youtu.be/ZysYNBI\_Y7w}}. We first obtain the original latent state $\phi(x_0)$ utilizing the original state $x_0$ which consists of $c$ adjacent images, then linearly predict $\phi _{\mathscr{K}}(x_{1:k})$ with a sequence of controls $u_{0:k-1}$ according to the recurrent rule in Fig. \ref{fig:DCKNet_constructure} (b). Results show identified dynamics can not only accurately predict dynamical intrinsic features, such as positions, angles, and velocities, but also contain fixed information of environments, i.e. the size of the pole, and the shape of the mountain. Generated pixels demonstrate that we do not approximate the Koopman modes in this work, but we still can realize full observations via the CNN decoder. \par
Consider the pixels error of generated images can not evaluate the identified performance precisely because small pixels error could correspond to a large real state error and vice versa. For instance, for the CartPole task, when the generated images have an accurate prediction at the angle feature, but there is a small bias at the position feature, this situation will show a large mean absolute error (MAE) on pixels. However, it is a satisfying prediction result for control. 
 To evaluate the performance of the latent dynamics, for intuitive visualization, we adopt the MAE of the latent states in episodes to evaluate the linear evolution performance of the identified latent dynamics:
\begin{equation}\label{equ:MAE}
\nonumber
MAE=\frac{1}{M}\sum{\phi _{\mathscr{K}}\left( x_{1:T} \right) -\phi \left( x_{1:T} \right)}
\end{equation}
where $M$ is the number of episodes to calculate the MAE, $T$ is the time steps of prediction. \par
\begin{figure*}[htbp]
	\centering
	\subfloat[MAEs of evolution on the latent states for CartPole]{
		\label{subfig:cp_error}
		\includegraphics[width=\textwidth]{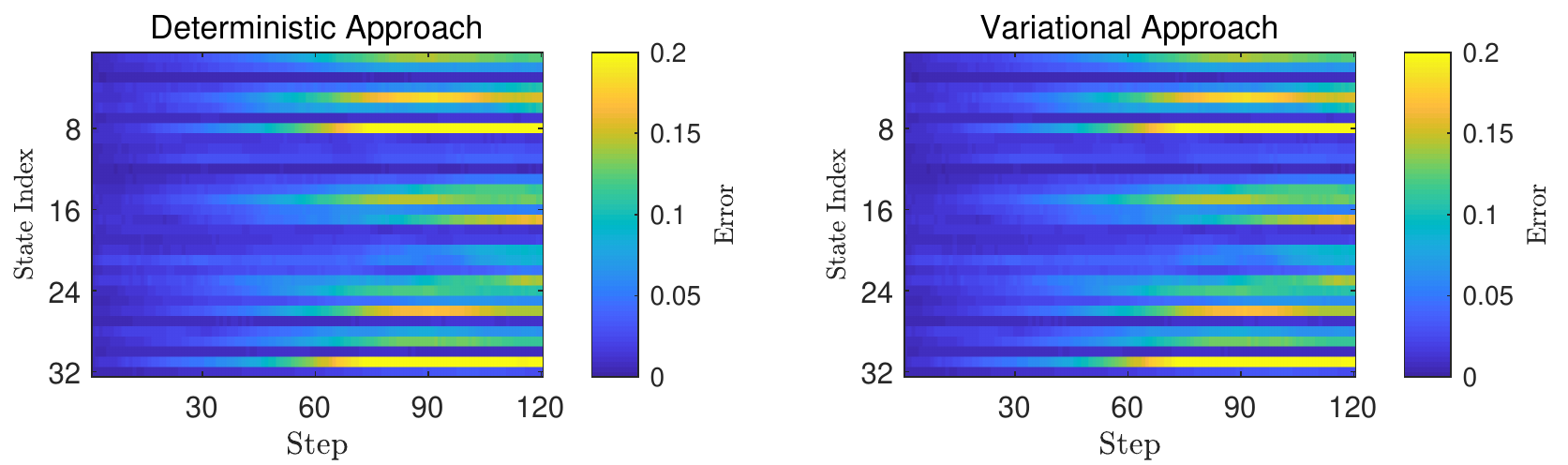}}\\
	\subfloat[MAEs of evolution on the latent states for MountainCar]{
		\label{subfig:vcp_error}
		\includegraphics[width=\textwidth]{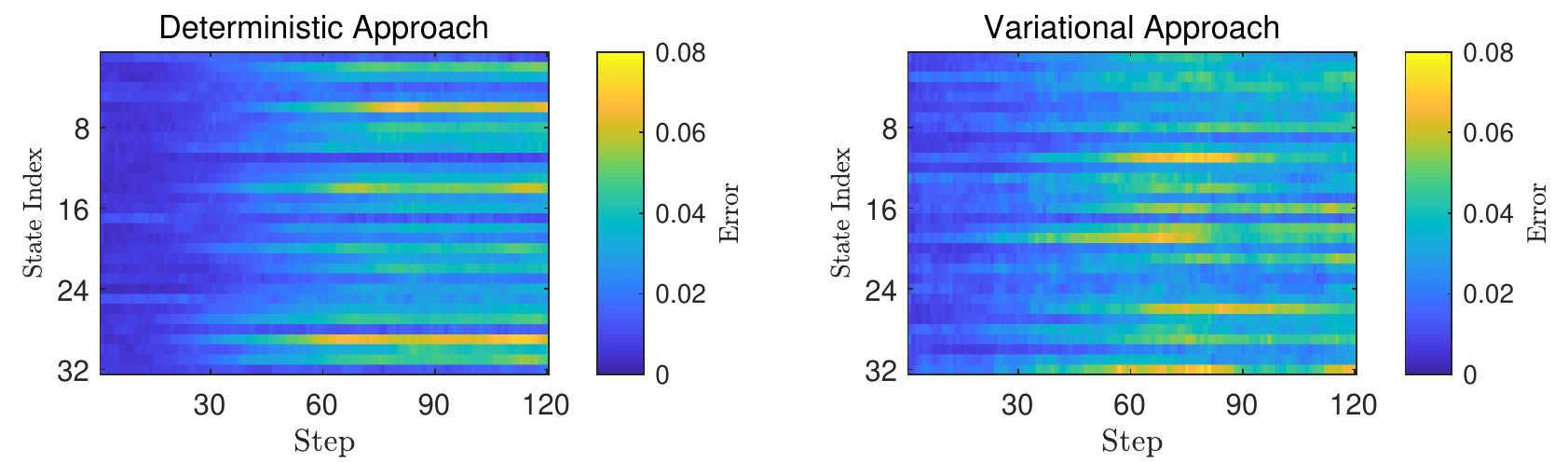}}\\
	\caption{MAEs of linear evolution on the latent states with DCKNet and VCKNet. For each approach of each task, we randomly sample 30 episodes from the testing dataset for calculating the MAEs. The left column denotes evolutionary MAEs of the CartPole and MountainCar tasks calculated by the DCKNet while the right column denotes MAEs calculated by the VCKNet. On the surface, the MAEs of evolution on the latent states of CartPole are much bigger than the MountainCar task, but as shown in Fig. \ref{fig:images_basisfunctions}, we can grasp the reason that the CartPole task has much bigger values of the latent states than the MountainCar task.}
	\label{fig:error_fig}
\end{figure*}
\begin{figure*}[htbp]
	\centering
	\subfloat[Evolving with the DCKNet for CartPole]{
		\label{subfig:cp_bf}
		\includegraphics[width=\textwidth]{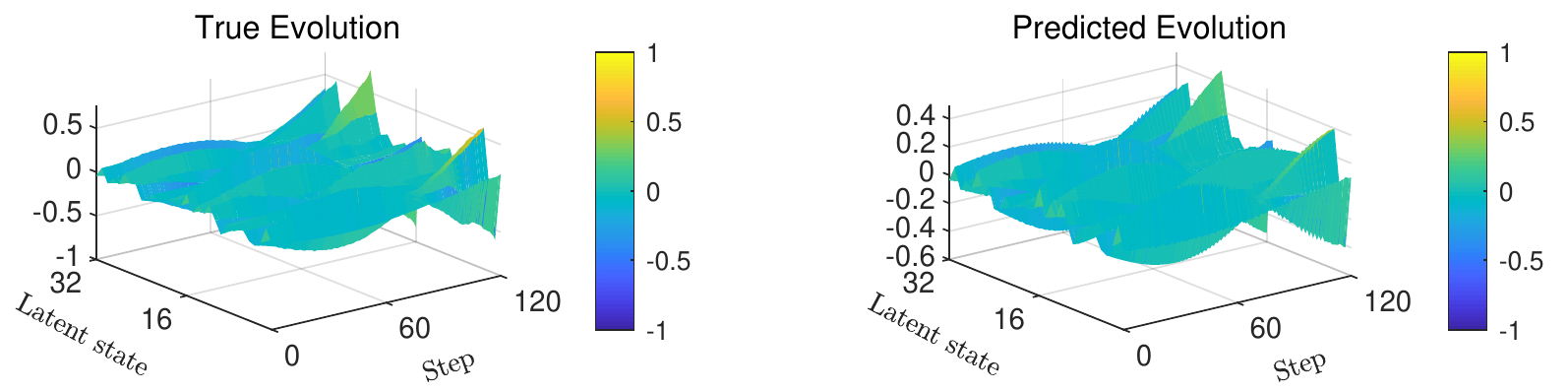}}\\
	\subfloat[Evolving with the VCKNet for CartPole]{
		\label{subfig:vcp_bf}
		\includegraphics[width=\textwidth]{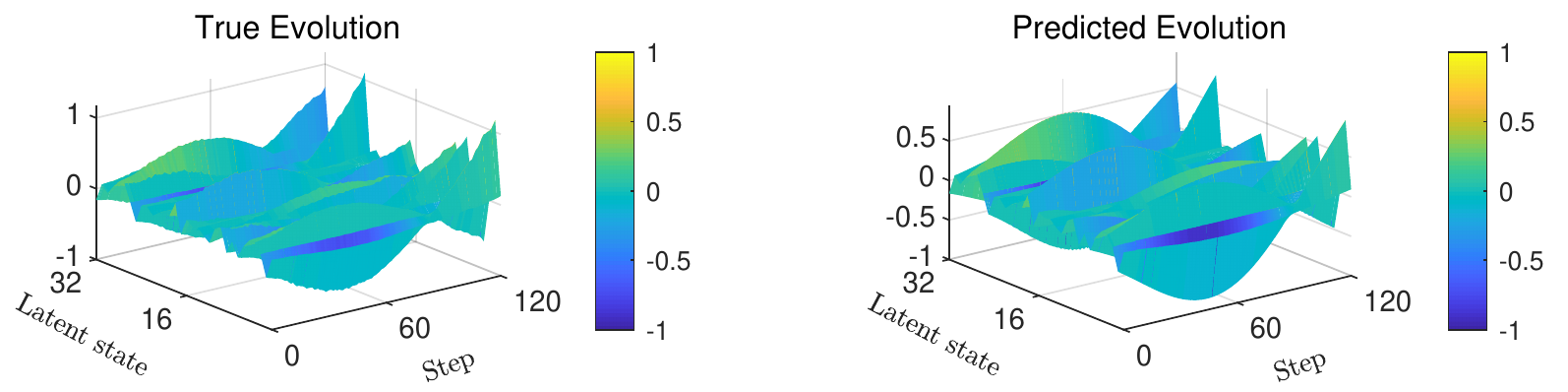}}\\
	\subfloat[Evolving with the DCKNet for MountainCar]{
		\label{subfig:mc_bf}
		\includegraphics[width=\textwidth]{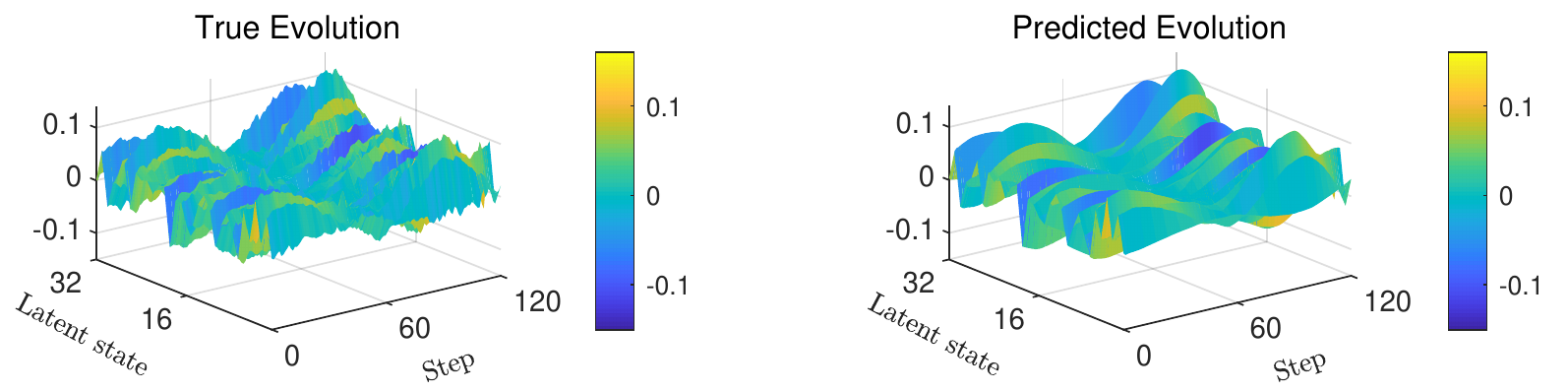}}\\
	\subfloat[Evolving with the VCKNet for MountainCar]{
		\label{subfig:vmc_bf}
		\includegraphics[width=\textwidth]{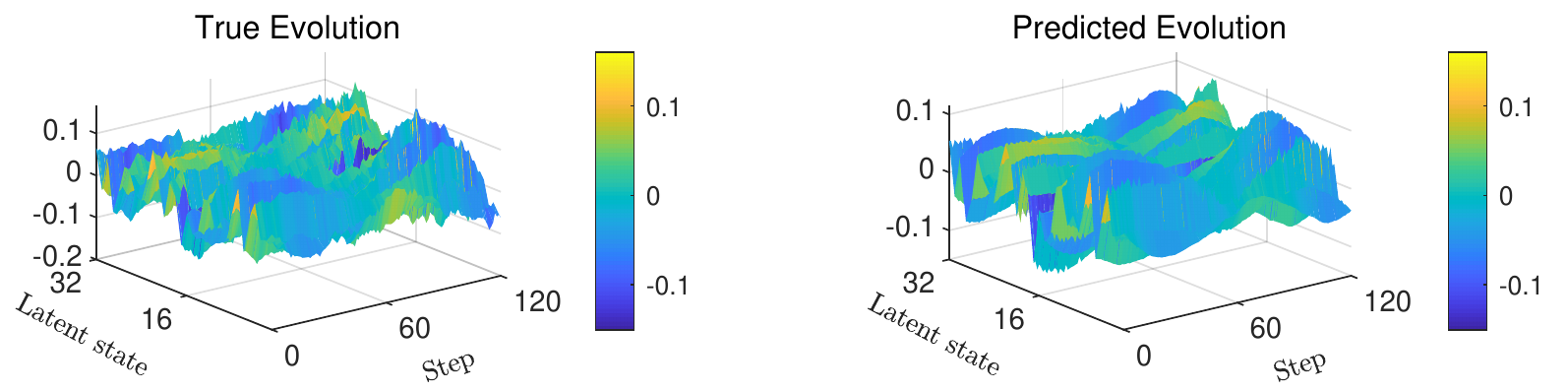}}\\
	\caption{Evolution example on the latent states of the CartPole and MountainCar tasks. (a, b) The same example episode of the CartPole task evolved with DCKNet and VCKNet separately. (c, d) Again, the same example episode of the MountainCar task evolved with DCKNet and VCKNet. For each task, we evolve the same episode with different approaches. The left column of the above subfigures is realized with the DCKNet while the right column denotes evolutionary episodes with the VCKNet. For the same episode of each task, the proposed approaches learn similar evolutionary distributions of the latent states.}
	\label{fig:images_basisfunctions}
\end{figure*}

As shown in Fig. \ref{fig:error_fig}, we sampled 30 episodes from the testing dataset for each task to calculate the evolutionary MAE of the latent states. It seems that learned models of the CartPole task have much bigger biases than the MountainCar task, but actually, the reason is the latent states of CartPole have a bigger valued distribution which we can know in Fig. \ref{fig:images_basisfunctions}. For both the CartPole and MountainCar tasks, DCKNet and VCKNet can obtain similar modeling performances that there are very small MAEs within 60 prediction steps. Most importantly, after 60 prediction steps, prediction MAEs stay in small ranges instead of divergence. And this phenomenon demonstrates these learned models approximate the Koopman operator of corresponding dynamics globally to some extent.
 \par

\textbf{Spectral analysis: }
The Koopman eigenvalues and corresponding eigenfunctions are intrinsic to dynamical systems. They play a key role in how the system evolves and responses. Based on the above simulation results, spectral analysis of the adopted two tasks with DCKNet and VCKNet are achieved respectively in this part. Because the appropriate numerical dimension to approximate the Koopman operator of systems is still an open issue, we are not enabled to determine the best approximation subspace and the true eigenvalues and their corresponding eigenfunctions of the adopted tasks. Under this circumstance, we randomly choose one episode for each task to analyze the evolution of the latent states and the Koopman eigenfunctions. \par
\begin{figure*}[htbp]
	\centering
	\subfloat[The approximated Koopman eigenvalues of the CartPole task]{
		\label{subfig:cp_eigenvalues}
		\includegraphics[width=\textwidth]{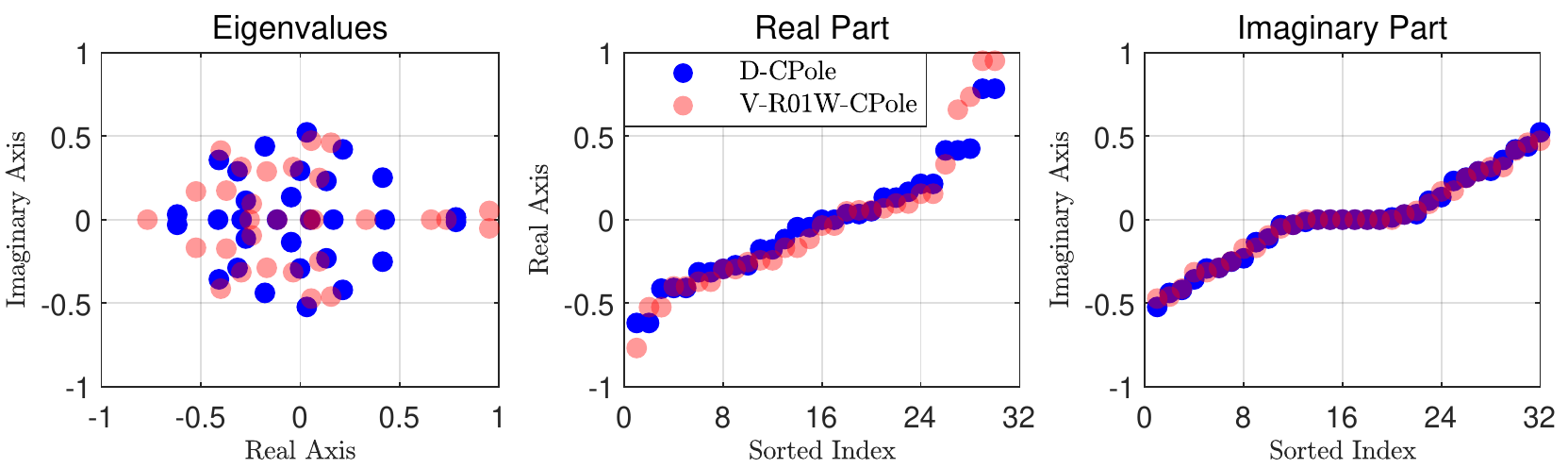}}\\
	\vspace{-0.1cm}
	\subfloat[The approximated Koopman eigenvalues of the MountainCar task]{
		\label{subfig:mc_eigenvalues}
		\includegraphics[width=\textwidth]{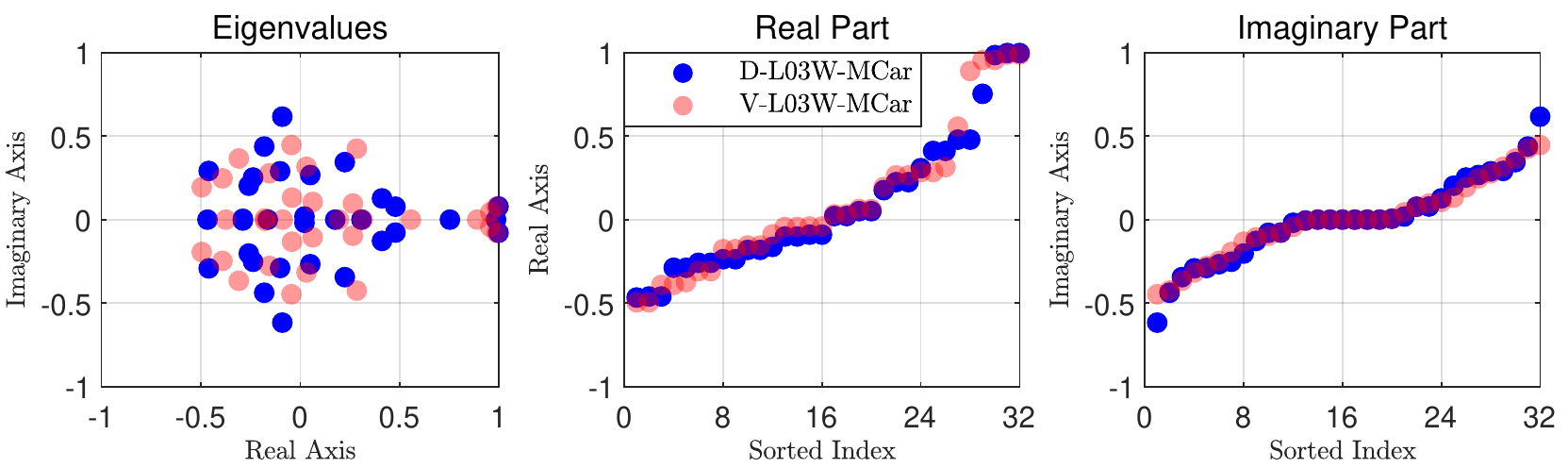}}\\
	\centering
	\caption{The Approximated eigenvalues of the adopted two tasks with DCKNet and VCKNet. The left figures of Fig. \ref{subfig:cp_eigenvalues} and Fig. \ref{subfig:mc_eigenvalues} are visualization of the eigenvalues of the state-transition matrix $A$ which are utilized to approximate the Koopman eigenvalues. Overall, eigenvalues of the same task with different approach have similar distributions. As the middle and right figures shown in Fig. \ref{subfig:cp_eigenvalues} and Fig. \ref{subfig:mc_eigenvalues}, we sort the eigenvalues in real and imaginary coordinates to visualize the distribution of eigenvalues more intuitively. Results show that different valid approaches learn similar distributions of the Koopman eigenvalues.}
	\vspace{-0.2cm}
	\label{fig:images_eigenvalues}
\end{figure*}

As shown in Fig. \ref{fig:images_eigenvalues}, eigenvalues of the state-transition matrix are also the approximated eigenvalues of the Koopman operator. For the same task, different approaches may learn different state-transition matrices, but the approximated Koopman eigenvalues of the learned model have similar distributions. Namely, the proposed DCKNet and VCKNet can model the intrinsic features of tasks from pixels. In fact, we discover the other learned models with different auxiliary weights also learn similar distributions of the Koopman eigenvalues. The real and imaginary parts of the Koopman eigenvalues are concentrated in the range of $[-1,\ 1]$, and eigenvalues are constant and not affected by the inputted pixels. These lead to the result that distributions of eigenvalues between different models that are not obvious. In addition, due to the Koopman eigenvalues denote the degrees of transforms in the directions of the Koopman eigenfunctions, eigenvalues are different in subspaces which are spanned by different eigenfunctions learned by different approaches. To present the evolving process of the latent dynamics more clearly, we visualize the latent states based on the same episode with both DCKNet and VCKNet. \par

As shown in Fig. \ref{fig:images_basisfunctions}, we randomly choose one episode for each task and execute the evolution on the latent states with the DCKNet and VCKNet separately. 
For the CartPole task, DCKNet and VCKNet obtain very similar evolutions in shapes except that the evolution obtained by the VCKNet has a bigger amplitude.
For the MountainCar task, DCKNet and VCKNet learn similar evolutions both in shapes and scales. Besides, predicted evolutions are smoother. 
It is noteworthy that though different approaches obtain similar performances (Fig. \ref{fig:images_result}) and evolutions, the learned basis functions are not in the same order, especially obvious in Fig. \ref{subfig:mc_bf} and Fig. \ref{subfig:vmc_bf}.
In this paper, basis functions induced by the CNN encoder are the states of the identified latent dynamics. It means that the evolution of the identified dynamics is limited in the space which is spanned by the basis functions. The evolution of the latent states reveals the dynamical shifting instead of the intrinsic features of the learned latent dynamics. \par

\begin{figure*}[htbp]
	\centering
	\subfloat[The evolution of the Koopman eigenfunctions of the CartPole task with the DCKNet]{
		\label{subfig:cp_DCK_eigenfunctions}
		\includegraphics[width=\textwidth]{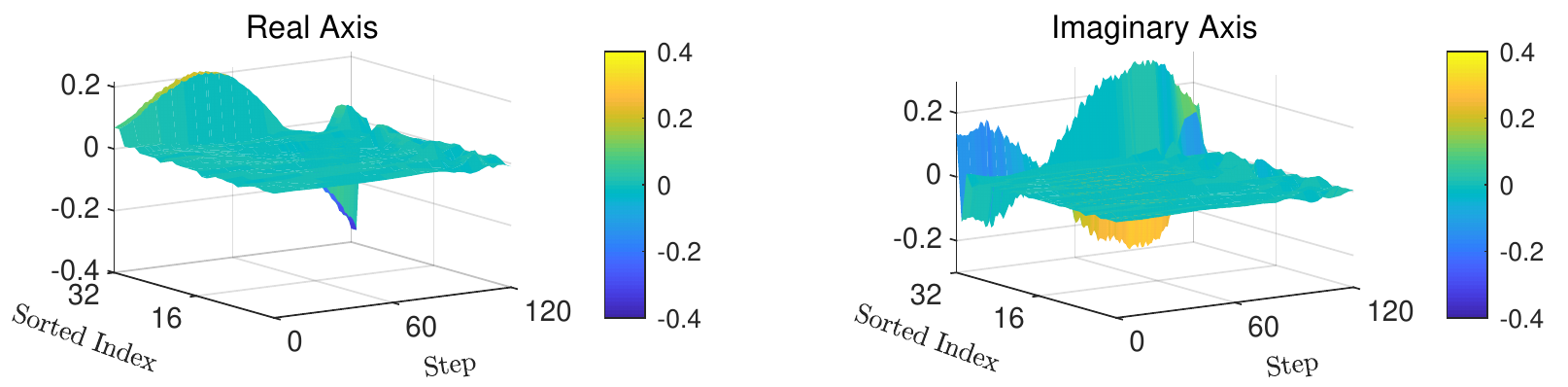}}\\
	\vspace{-0.1cm}
	\subfloat[The evolution of the Koopman eigenfunctions of the CartPole task the VCKNet]{
		\label{subfig:cp_VCK_eigenfunctions}
		\includegraphics[width=\textwidth]{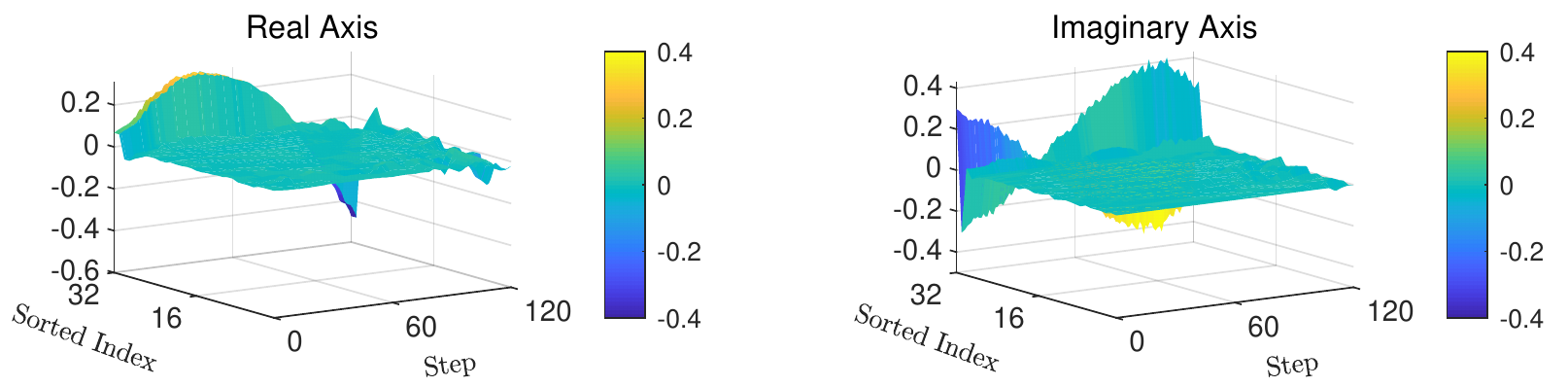}}\\
	\subfloat[The evolution of the Koopman eigenfunctions of the MountainCar task with the deterministic approach]{
		\label{subfig:mc_DCK_eigenfunctions}
		\includegraphics[width=\textwidth]{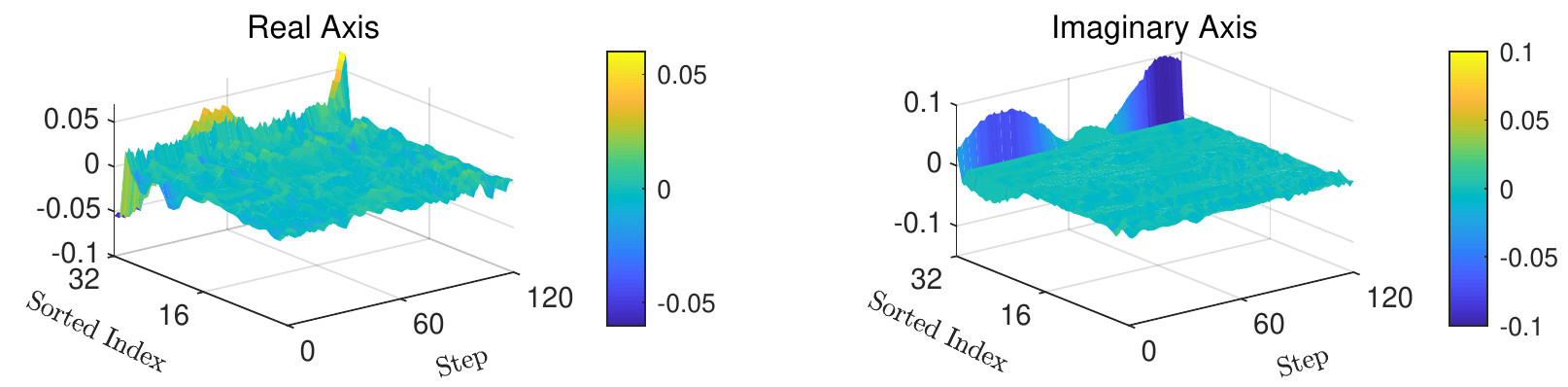}}\\
	\vspace{-0.1cm}
	\subfloat[The evolution of the Koopman eigenfunctions of the MountainCar task the variational approach]{
		\label{subfig:mc_VCK_eigenfunctions}
		\includegraphics[width=\textwidth]{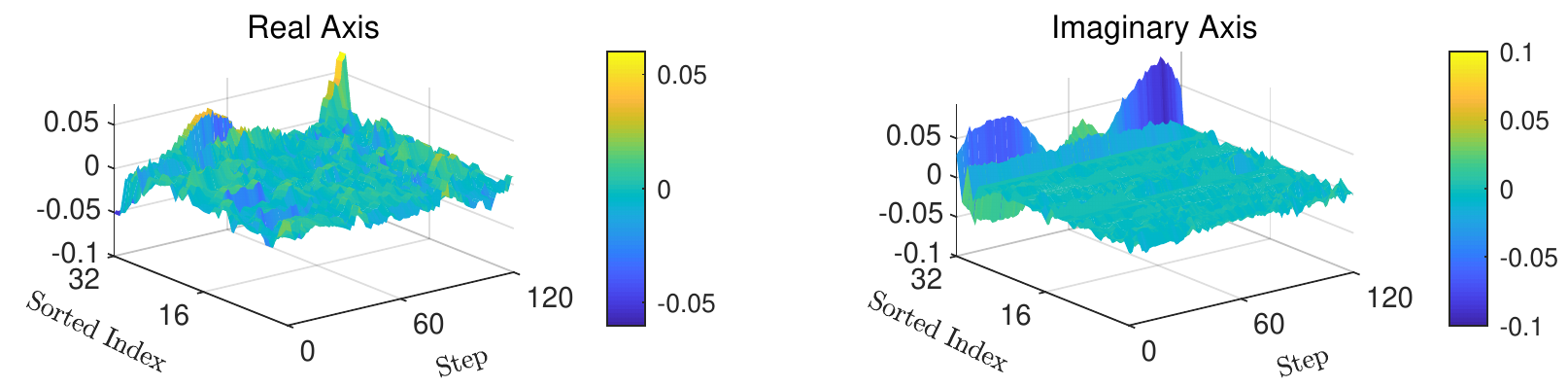}}\\
	\centering
	\caption{The evolution of the approximated Koopman eigenfunctions via different approaches. The above evolutionary Koopman eigenfunctions are calculated based on the same episode corresponding to tasks shown in Fig. \ref{fig:images_basisfunctions}. For the CartPole task, DCKNet and VCKNet learned similar evolutions of the Koopman eigenfunctions in shapes both the real or imaginary parts except. From Fig. \ref{fig:images_basisfunctions}, we can know that the differences in shapes are caused by the latent state. For the MountainCar task, CKnet learns very similar evolutions both in shapes and scales.}
	\vspace{-0.2cm}
	\label{fig:images_eigenfunctions}
\end{figure*}

Except for the Koopman eigenvalues, the Koopman eigenfunctions are also the intrinsic features of a system. Different from the Koopman eigenvalues, the Koopman eigenfunctions are changing according to the current latent states. In order to visualize the intrinsic characteristics more intuitively, the Koopman eigenfunctions are shown in Fig. \ref{fig:images_eigenfunctions}. Recall that the Koopman eigenfunctions are calculated by $\varphi = \varXi ^{\top} \phi$. Because different approaches do not learn the same order of the basis functions $\phi$ and a large eigenvalue means more information in the corresponding dimension, we sort the Koopman eigenvalues according and reorder the eigenvector based on the sorting indexes for calculating the Koopman eigenfunctions. \par
Evolutionary results of the Koopman eigenfunctions are similar to the latent states that evolutions have similar shapes both in real and imaginary parts. Eigenvalues denote the magnitude of transformations in directions of their associated eigenvectors. Therefore, for the CartPole task shown in Fig. \ref{subfig:cp_DCK_eigenfunctions} and Fig. \ref{subfig:cp_VCK_eigenfunctions}, the Koopman eigenfunctions with different approaches have approximately double gap in amplitude both in real and imaginary parts. This does not mean the VCKNet learns more information on directions of the basis vectors but magnifies the same size in each direction. For the MountainCar task, the proposed two approaches obtain more similar evolutions based on the same episode. 
In overall, the proposed approaches can learn effective models, though the evolutionary shapes of the Koopman eigenfunctions have slight differences. This phenomenon is normal because models learned by different approaches have unequal state transitions leading to irrelative subspaces, and the VCKNet acquires the latent states by sampling from the learned Gaussian distribution leading to biases to the eigenfunctions. Besides, note that the above Koopman eigenfunctions are corresponding to the sorted eigenvalues, the results show that eigenfunctions have smaller fluctuations in evolutionary processes if the corresponding eigenvalues are smaller.

\subsection{Cases in Mujoco}
\begin{figure*}[htbp]
	\centering
	\subfloat[Prediction results visualization of the Cheetah-run task]{
		\label{subfig:mujoco_cheetah_run}
		\includegraphics[scale=0.4]{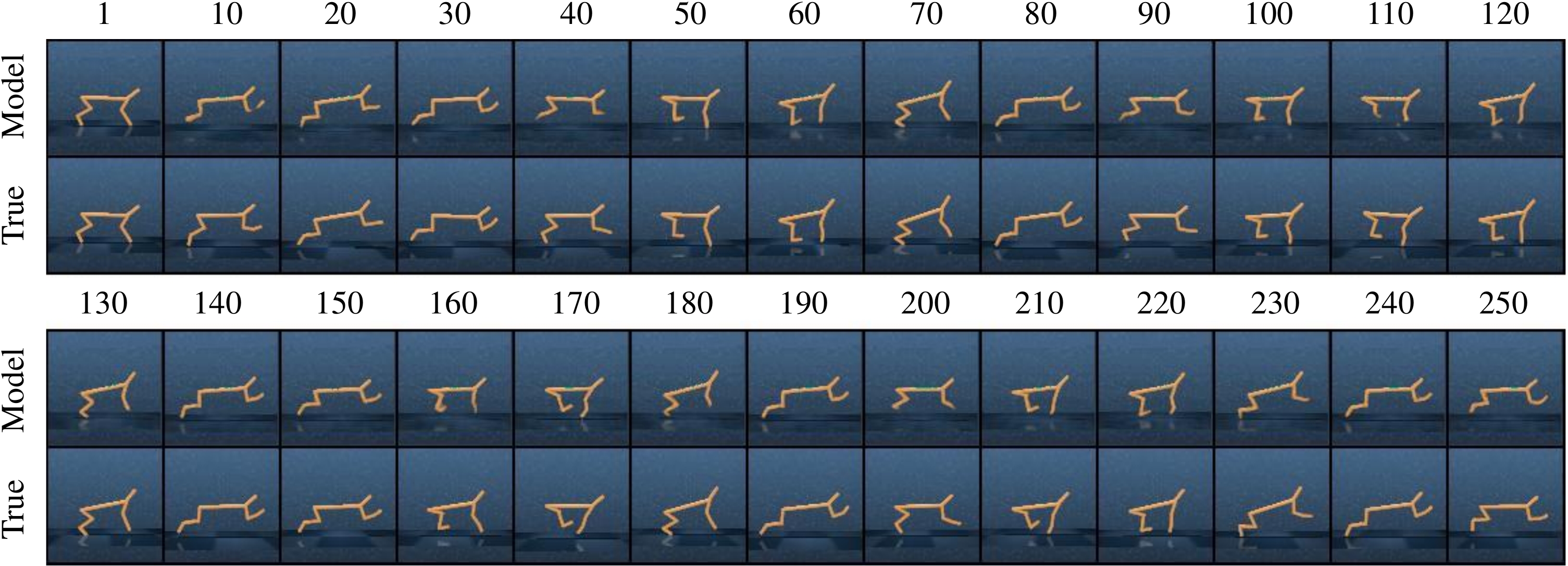}}\\
	\vspace{-0.1cm}
	\subfloat[Prediction results visualization of the Ball\_in\_cup-catch task]{
		\label{subfig:mujoco_ball_in_cup_catch}
		\includegraphics[scale=0.4]{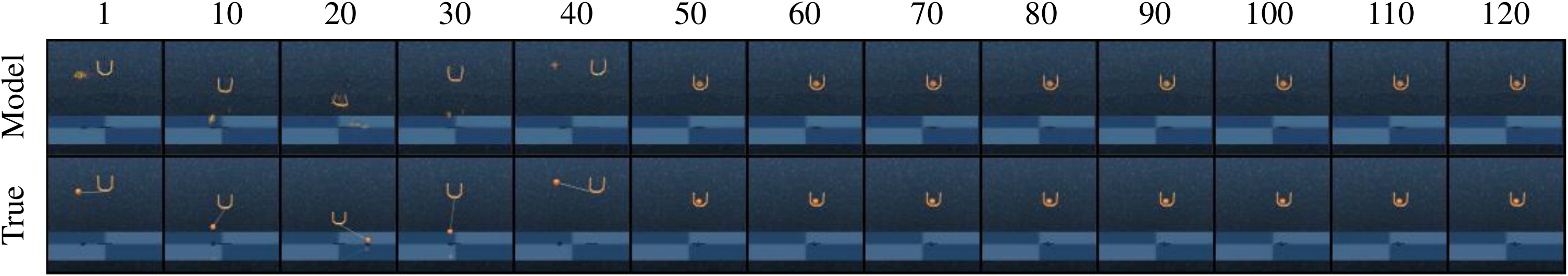}}\\
	\subfloat[Prediction results visualization of the Cartpole-balance task]{
		\label{subfig:mujoco_cartpole_balance}
		\includegraphics[scale=0.4]{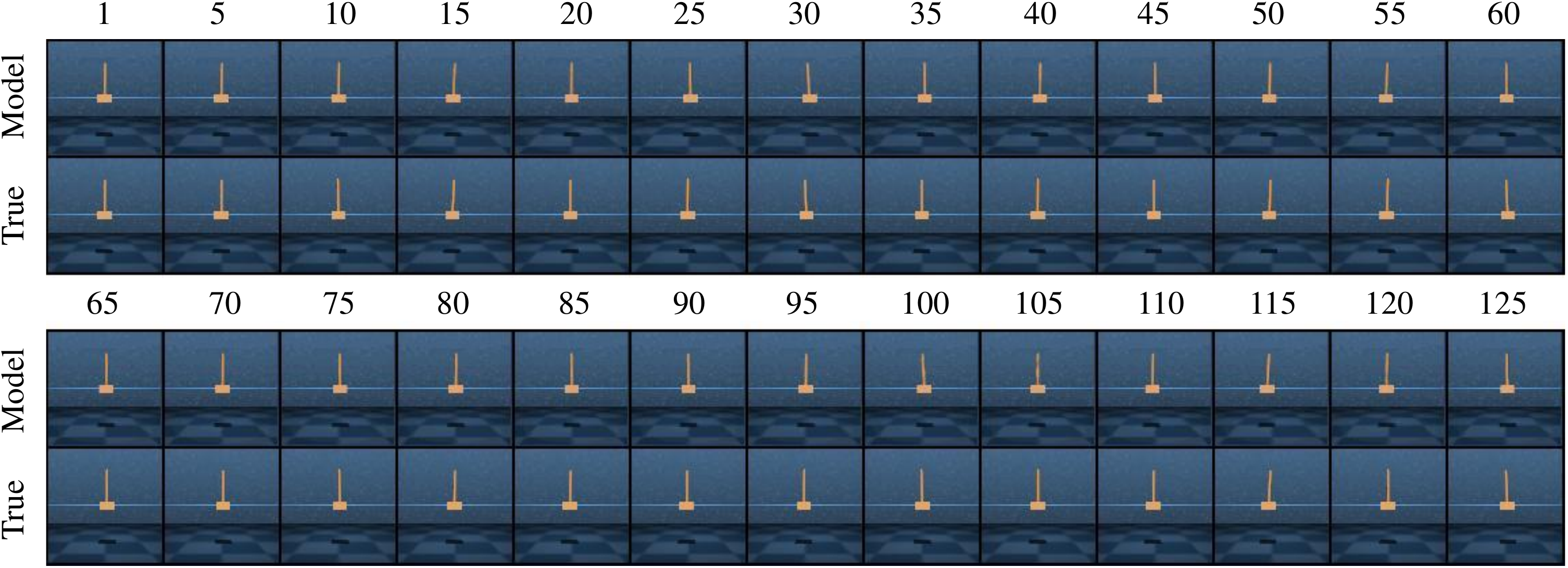}}\\
	\subfloat[Prediction results visualization of the Walker-walk task]{
		\label{subfig:mujoco_walker_walk}
		\includegraphics[scale=0.4]{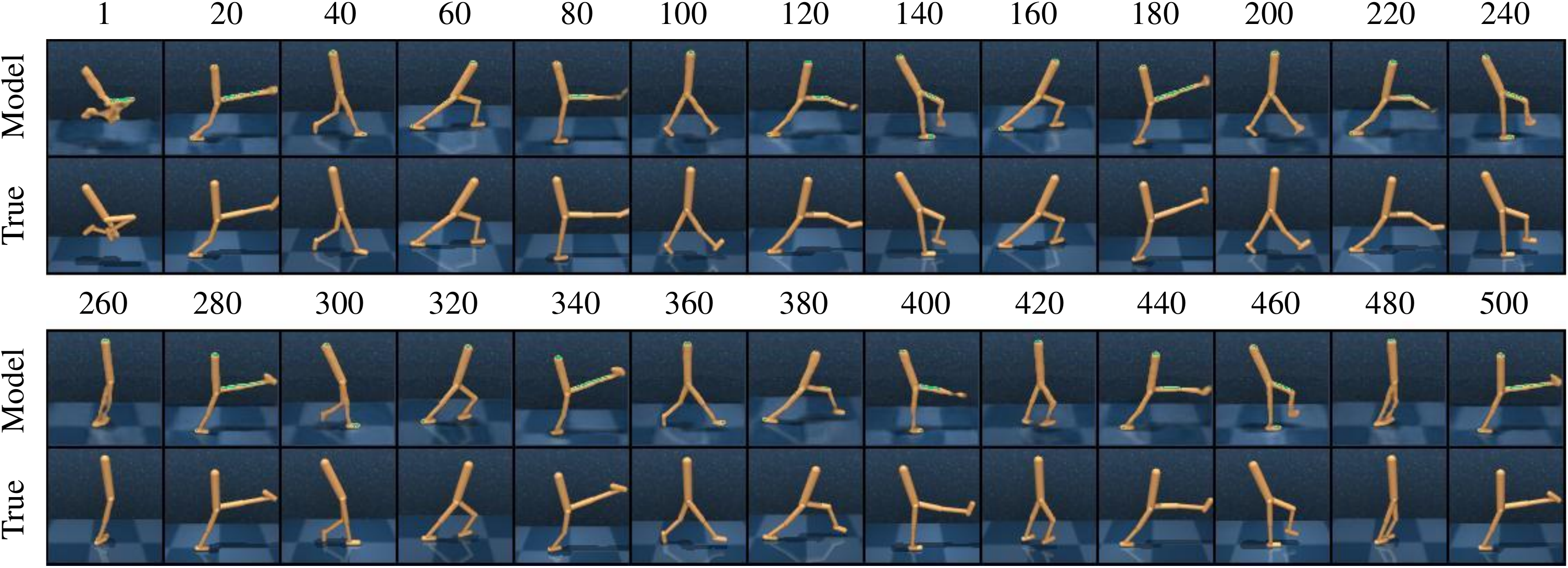}}\\
	\centering
	\caption{Koopman evolution prediction results of Mujoco cases. Each task runs 1000 steps in Mujoco, but the action repeat trick is usually adopted to improve performances of DRL.}
	\vspace{-0.2cm}
	\label{fig:mujoco_images_result}
\end{figure*}

\textbf{Identification accuracy analysis: }
Linear evolution prediction result are shown in Fig. \ref{fig:mujoco_images_result}. Mujoco cases are trained in online manner with SAC, and a detailed prediction video about these four cases is publicly available\footnote{\href{https://www.youtube.com/watch?v=5TKdlesbHvc}{https://www.youtube.com/watch?v=5TKdlesbHvc}}. For each task in Mujoco, we raise $1000/A_r$ prediction steps, i.e. for Cheetah-run and Ball\_in\_cup-catch, prediction steps equal 250 because $A_r$ equals 4. Note that we did not draw images after 120 steps because images are constant after that.
\par
\begin{figure*}[htbp]
	\begin{center}
		\includegraphics[scale=0.85]{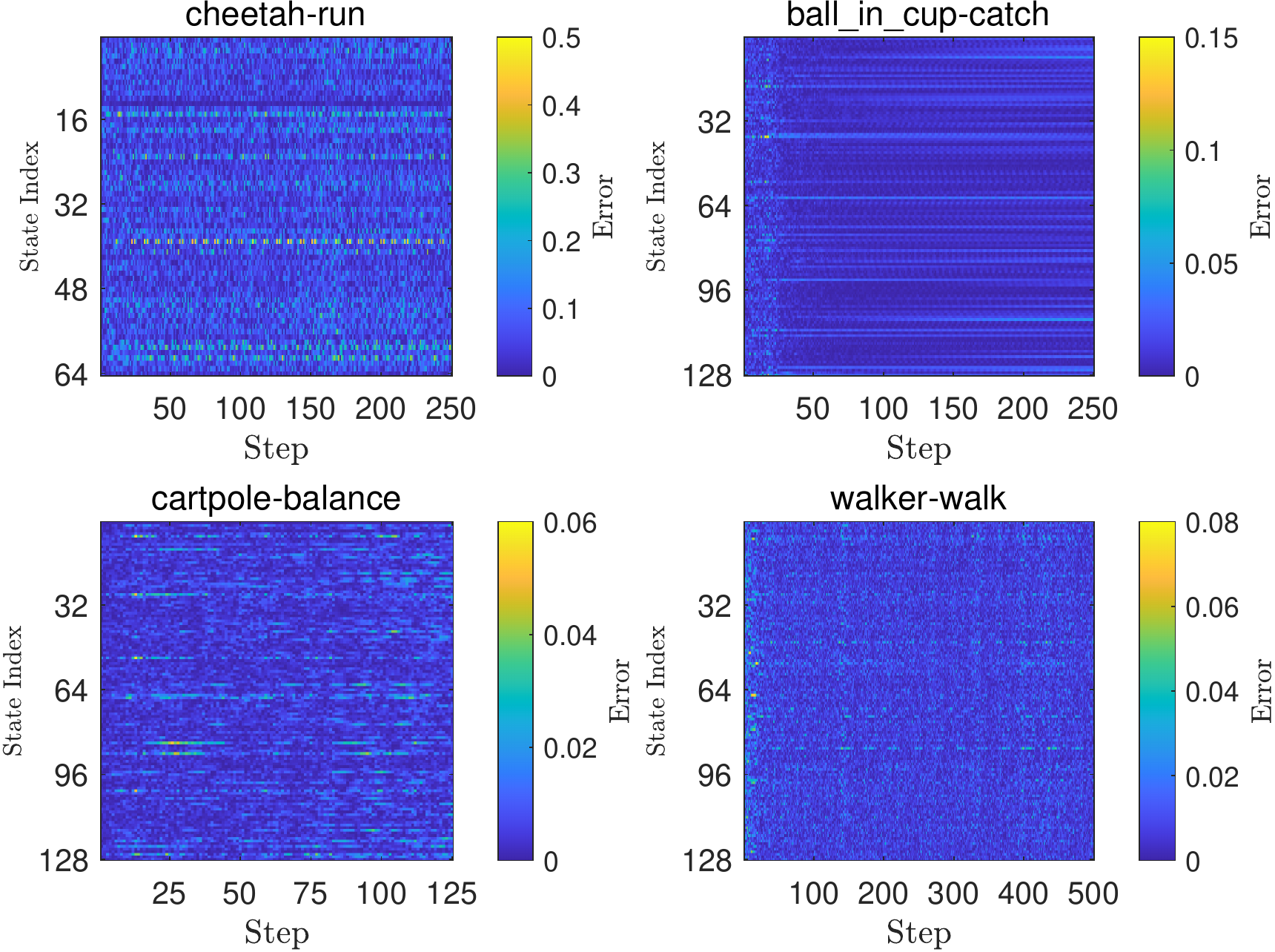}
		\caption{MAEs of linear evolution on the latent states for different cases.}
		\label{fig:maes_mujoco}
	\end{center}
\end{figure*}

Except for the prediction of images, we pay more attention to linear evolution on latent states because we usually utilize latent states for control, prediction, etc. Fig. \ref{fig:maes_mujoco} shows linear evolution MAEs of latent states for the adopted Mujoco cases. In line with performances in Fig. \ref{fig:mujoco_images_result}, ball\_in\_cap-catch and walker-walk have greater errors in the early stage while cheetah-run and cartpole-balance perform excellently throughout the whole prediction stages.

Combining the results in Fig. \ref{subfig:mujoco_ball_in_cup_catch} and Fig. \ref{subfig:mc_result}, We are aware that CKNet is not suitable for cases that objects are small relative to inputted image sizes and move quickly.

	\section{Conclusions}\label{sec:conclusions}
	This paper proposes a novel Koopman-based deep learning approach called CKNet for identifying latent dynamics from pixels. CKNet is realized with two different approaches, the DCKNet and VCKNet. Besides, auxiliary weights are introduced into the multi-step linearity and prediction losses to improve the prediction performance. Since the training process is designed under the constraints of the Koopman operator, the identified model is linear, controllable, and physically interpretable in the subspace constructed by the encoder. Experiments were conducted on two offline trained and four online trained nonlinear forced dynamical systems with continuous action spaces in Gym and Mujoco, and the results show that identified models can accurately predict the latent states for long-term steps without divergence and generate corresponding clear images. In particular, we analyze the evolutionary processes of the latent states and the Koopman eigenfunctions based on the same episode with the DCKNet and VCKnet separately in Gym cases and the results confirm that CKNet learns similar evolutions of the Koopman eigenfunctions in shapes whether in the form of DCKNet or VCKNet. Directions for future work include paying more attention to systems that can only be described with pixels, such as Atari games and autonomous driving. Besides, RL algorithms based on CKNet will also be researched in the future.
	\ifCLASSOPTIONcaptionsoff
	\newpage
	\fi
	
    \normalem
	\bibliographystyle{IEEEtran}
	\bibliography{CKNet,IEEEabrv}

\begin{thebibliography}{10}
\providecommand{\url}[1]{#1}
\csname url@samestyle\endcsname
\providecommand{\newblock}{\relax}
\providecommand{\bibinfo}[2]{#2}
\providecommand{\BIBentrySTDinterwordspacing}{\spaceskip=0pt\relax}
\providecommand{\BIBentryALTinterwordstretchfactor}{4}
\providecommand{\BIBentryALTinterwordspacing}{\spaceskip=\fontdimen2\font plus
\BIBentryALTinterwordstretchfactor\fontdimen3\font minus
  \fontdimen4\font\relax}
\providecommand{\BIBforeignlanguage}[2]{{%
\expandafter\ifx\csname l@#1\endcsname\relax
\typeout{** WARNING: IEEEtran.bst: No hyphenation pattern has been}%
\typeout{** loaded for the language `#1'. Using the pattern for}%
\typeout{** the default language instead.}%
\else
\language=\csname l@#1\endcsname
\fi
#2}}
\providecommand{\BIBdecl}{\relax}
\BIBdecl

\bibitem{Christof2017On}
C.~Schütte, P.~Koltai, and S.~Klus, ``On the numerical approximation of the
  perron-frobenius and koopman operator,'' \emph{Journal of Computational
  Dynamics}, vol.~3, no.~1, pp. 51--79, 2017.

\bibitem{2017On}
M.~Korda and I.~Mezic, ``On convergence of extended dynamic mode decomposition
  to the koopman operator,'' \emph{Journal of Nonlinear Science}, vol.~28, 04
  2018.

\bibitem{schmid2010dynamic}
P.~J. Schmid, ``Dynamic mode decomposition of numerical and experimental
  data,'' \emph{Journal of fluid mechanics}, vol. 656, pp. 5--28, 2010.

\bibitem{tu2013dynamic}
J.~H. Tu, C.~W. Rowley, D.~M. Luchtenburg, S.~L. Brunton, and J.~N. Kutz, ``On
  dynamic mode decomposition: Theory and applications,'' \emph{arXiv preprint
  arXiv:1312.0041}, 2013.

\bibitem{arbabi2017ergodic}
H.~Arbabi and I.~Mezic, ``Ergodic theory, dynamic mode decomposition, and
  computation of spectral properties of the koopman operator,'' \emph{SIAM
  Journal on Applied Dynamical Systems}, vol.~16, no.~4, pp. 2096--2126, 2017.

\bibitem{brunton2017chaos}
S.~L. Brunton, B.~W. Brunton, J.~L. Proctor, E.~Kaiser, and J.~N. Kutz, ``Chaos
  as an intermittently forced linear system,'' \emph{Nature communications},
  vol.~8, no.~1, pp. 1--9, 2017.

\bibitem{hemati2014dynamic}
M.~S. Hemati, M.~O. Williams, and C.~W. Rowley, ``Dynamic mode decomposition
  for large and streaming datasets,'' \emph{Physics of Fluids}, vol.~26,
  no.~11, p. 111701, 2014.

\bibitem{kevrekidis2016kernel}
I.~G. Kevrekidis, C.~W. Rowley, and M.~O. Williams, ``A kernel-based method for
  data-driven koopman spectral analysis,'' \emph{Journal of Computational
  Dynamics}, vol.~2, no.~2, pp. 247--265, 2016.

\bibitem{avila2020data}
A.~Avila and I.~Mezi{\'c}, ``Data-driven analysis and forecasting of highway
  traffic dynamics,'' \emph{Nature communications}, vol.~11, no.~1, pp. 1--16,
  2020.

\bibitem{williams2015data}
M.~O. Williams, I.~G. Kevrekidis, and C.~W. Rowley, ``A data--driven
  approximation of the koopman operator: Extending dynamic mode
  decomposition,'' \emph{Journal of Nonlinear Science}, vol.~25, no.~6, pp.
  1307--1346, 2015.

\bibitem{mezic2013analysis}
I.~Mezi{\'c}, ``Analysis of fluid flows via spectral properties of the koopman
  operator,'' \emph{Annual Review of Fluid Mechanics}, vol.~45, pp. 357--378,
  2013.

\bibitem{susuki2016applied}
Y.~Susuki, I.~Mezic, F.~Raak, and T.~Hikihara, ``Applied koopman operator
  theory for power systems technology,'' \emph{Nonlinear Theory and Its
  Applications, IEICE}, vol.~7, no.~4, pp. 430--459, 2016.

\bibitem{netto2018robust}
M.~Netto and L.~Mili, ``A robust data-driven koopman kalman filter for power
  systems dynamic state estimation,'' \emph{IEEE Transactions on Power
  Systems}, vol.~33, no.~6, pp. 7228--7237, 2018.

\bibitem{klus2018kernel}
S.~Klus, A.~Bittracher, I.~Schuster, and C.~Sch{\"u}tte, ``A kernel-based
  approach to molecular conformation analysis,'' \emph{The Journal of Chemical
  Physics}, vol. 149, no.~24, p. 244109, 2018.

\bibitem{mamakoukas2019local}
G.~Mamakoukas, M.~L. Castano, X.~Tan, and T.~Murphey, ``Local koopman operators
  for data-driven control of robotic systems.'' in \emph{Robotics: science and
  systems}, 2019.

\bibitem{proctor2018generalizing}
J.~L. Proctor, S.~L. Brunton, and J.~N. Kutz, ``Dynamic mode decomposition with
  control,'' \emph{SIAM Journal on Applied Dynamical Systems}, vol.~17, no.~1,
  pp. 142--161, 2018.

\bibitem{williams2016extending}
M.~O. Williams, M.~S. Hemati, S.~T. Dawson, I.~G. Kevrekidis, and C.~W. Rowley,
  ``Extending data-driven koopman analysis to actuated systems,''
  \emph{IFAC-PapersOnLine}, vol.~49, no.~18, pp. 704--709, 2016.

\bibitem{korda2018linear}
M.~Korda and I.~Mezi{\'c}, ``Linear predictors for nonlinear dynamical systems:
  Koopman operator meets model predictive control,'' \emph{Automatica},
  vol.~93, pp. 149--160, 2018.

\bibitem{morton2018deep}
J.~Morton, F.~D. Witherden, A.~Jameson, and M.~J. Kochenderfer, ``Deep
  dynamical modeling and control of unsteady fluid flows,'' \emph{arXiv
  preprint arXiv:1805.07472}, 2018.

\bibitem{ping2020deep}
Z.~Ping, Z.~Yin, X.~Li, Y.~Liu, and T.~Yang, ``Deep koopman model predictive
  control for enhancing transient stability in power grids,''
  \emph{International Journal of Robust and Nonlinear Control}, 2020.

\bibitem{xiao2020deep}
Y.~Xiao, X.~Zhang, X.~Xu, X.~Liu, and J.~Liu, ``A deep learning framework based
  on koopman operator for data-driven modeling of vehicle dynamics,''
  \emph{arXiv preprint arXiv:2007.02219}, 2020.

\bibitem{mardt2018vampnets}
A.~Mardt, L.~Pasquali, H.~Wu, and F.~No{\'e}, ``Vampnets for deep learning of
  molecular kinetics,'' \emph{Nature communications}, vol.~9, no.~1, pp. 1--11,
  2018.

\bibitem{xie2019graph}
T.~Xie, A.~France-Lanord, Y.~Wang, Y.~Shao-Horn, and J.~C. Grossman, ``Graph
  dynamical networks for unsupervised learning of atomic scale dynamics in
  materials,'' \emph{Nature communications}, vol.~10, no.~1, pp. 1--9, 2019.

\bibitem{morton2019Deep}
J.~Morton, F.~D. Witherden, and M.~J. Kochenderfer, ``Deep variational koopman
  models: Inferring koopman observations for uncertainty-aware dynamics
  modeling and control,'' in \emph{Twenty-Eighth International Joint Conference
  on Artificial Intelligence IJCAI-19}, 2019.

\bibitem{otto2019linearly}
S.~E. Otto and C.~W. Rowley, ``Linearly recurrent autoencoder networks for
  learning dynamics,'' \emph{SIAM Journal on Applied Dynamical Systems},
  vol.~18, no.~1, pp. 558--593, 2019.

\bibitem{lusch2018deep}
B.~Lusch, J.~N. Kutz, and S.~L. Brunton, ``Deep learning for universal linear
  embeddings of nonlinear dynamics,'' \emph{Nature communications}, vol.~9,
  no.~1, pp. 1--10, 2018.

\bibitem{ostafew2016learning}
C.~J. Ostafew, A.~P. Schoellig, T.~D. Barfoot, and J.~Collier, ``Learning-based
  nonlinear model predictive control to improve vision-based mobile robot path
  tracking,'' \emph{Journal of Field Robotics}, vol.~33, no.~1, pp. 133--152,
  2016.

\bibitem{mnih2015human}
V.~Mnih, K.~Kavukcuoglu, D.~Silver, A.~A. Rusu, J.~Veness, M.~G. Bellemare,
  A.~Graves, M.~Riedmiller, A.~K. Fidjeland, G.~Ostrovski \emph{et~al.},
  ``Human-level control through deep reinforcement learning,'' \emph{nature},
  vol. 518, no. 7540, pp. 529--533, 2015.

\bibitem{haarnoja2018soft}
T.~Haarnoja, A.~Zhou, P.~Abbeel, and S.~Levine, ``Soft actor-critic: Off-policy
  maximum entropy deep reinforcement learning with a stochastic actor,'' in
  \emph{International Conference on Machine Learning}.\hskip 1em plus 0.5em
  minus 0.4em\relax PMLR, 2018, pp. 1861--1870.

\bibitem{schrittwieser2020mastering}
J.~Schrittwieser, I.~Antonoglou, T.~Hubert, K.~Simonyan, L.~Sifre, S.~Schmitt,
  A.~Guez, E.~Lockhart, D.~Hassabis, T.~Graepel \emph{et~al.}, ``Mastering
  atari, go, chess and shogi by planning with a learned model,'' \emph{Nature},
  vol. 588, no. 7839, pp. 604--609, 2020.

\bibitem{laskin2020curl}
M.~Laskin, A.~Srinivas, and P.~Abbeel, ``Curl: Contrastive unsupervised
  representations for reinforcement learning,'' in \emph{International
  Conference on Machine Learning}.\hskip 1em plus 0.5em minus 0.4em\relax PMLR,
  2020, pp. 5639--5650.

\bibitem{hafner2019planet}
D.~Hafner, T.~Lillicrap, I.~Fischer, R.~Villegas, D.~Ha, H.~Lee, and
  J.~Davidson, ``Learning latent dynamics for planning from pixels,'' in
  \emph{International Conference on Machine Learning}, 2019, pp. 2555--2565.

\bibitem{haq2019universal}
I.~U. Haq and Y.~Kawahara, ``Universal modal embedding of dynamics in videos
  and its applications,'' 2019.

\bibitem{van2020deepkoco}
B.~van~der Heijden, L.~Ferranti, J.~Kober, and R.~Babuska, ``Deepkoco:
  Efficient latent planning with an invariant koopman representation,''
  \emph{arXiv preprint arXiv:2011.12690}, 2020.

\bibitem{brockman2016openai}
G.~Brockman, V.~Cheung, L.~Pettersson, J.~Schneider, J.~Schulman, J.~Tang, and
  W.~Zaremba, ``Openai gym,'' \emph{arXiv preprint arXiv:1606.01540}, 2016.

\end{thebibliography}
	\newpage
	\onecolumn
	\begin{appendices} 
		\renewcommand\thefigure{\Alph{section}\arabic{figure}} 
		\section{Analytic dynamics}\label{sec:analytic_dynamics}
		\begin{equation}
			\begin{cases}
				\dot{x}=p_1u-p_2\cos \left( 3x \right)&		\dot{x}\in \left[ \dot{x}_{\min}\,\,\dot{x}_{\max} \right]\\
				\dot{x}=0&		x=x_{\min},\dot{x}<0\\
				\dot{x}=\dot{x}_{min}&		\dot{x}<\dot{x}_{min}\\
				\dot{x}=\dot{x}_{max}&		\dot{x}>\dot{x}_{max}\\
			\end{cases}
		\end{equation}
		where $p_1$ and $p_2$ are parameters of the system and they equal $0.0015$ and $0.0025$ respectively, $x$ denotes the car's position in horizon. As a result, the position of discrete-time for MountainCar is given as:
		\begin{equation}
			x = x + \dot{x} \Delta t
		\end{equation}
		where $\Delta t$ is the sample time which equals to $1s$. \par
		However for the CartPole task, Gym only supports discrete action space of $\left\{ -10,0,10 \right\} $, we utilize another version with acceleration as the control to support continuous action space for the CartPole task. The model's angular acceleration is given as follows:
		\begin{equation}
			\ddot{\theta}=\frac{3\Delta t}{4l}\left( g\sin \theta +u\cos \theta \right) 
		\end{equation}
		where $l$ is the length of the pole's center of gravity to the car, $g$ is the gravitational acceleration, $\theta$ indicates the angle between the pole and the vertical, $u$ is the acceleration of the car. Again, $\Delta t$ is the sample time which equals $0.02s$. Other state values can be calculated as follows:
		\begin{equation}
			\mathfrak{X}=\mathfrak{X}+\dot{\mathfrak{X}} \Delta t
		\end{equation}
		where $\mathfrak{X} \in \{x,\ \dot{x},\ \theta,\ \dot{\theta} \}$ denotes the car's position and velocity, the pole's angle and angular velocity respectively.  \par
		
		\section{Cases in Mujoco}
		\setcounter{figure}{0} 		
		
		The training process of CKNet with SAC are shown in Algorithm. \ref{alg:CK-SAC}, where $\theta_{sa}$ denotes weights of actor neural networks while $\theta_{sc}$ is weights of critic neural networks that include weights of the encoder $\theta_e$ and the critic $\theta_c$, $\theta_K$ indicates weights of the Koopman operator that includes matrices of $A$ and $B$, weights of the encoder $\theta_e$ and decoder $\theta_d$. Note that the Koopman operator shares the weights of the encoder and decoder with SAC. $\theta_{sa}$ denotes weights of the actor neural network $\theta_a$ and the temperature parameter $\aleph$. $J_{V}$ and $J_{\pi}$ are the losses of the value function and expected KL-divergence. $L$ is the total loss function of CKNet in \eqref{equ:total_loss}. $\beta_{\star}$ means the learning rate to the weights $\star$. More details of SAC could refer to \cite{haarnoja2018soft}. 
		\begin{algorithm}[h]
			\footnotesize
			\caption{The CK-SAC Algorithm}
			\label{alg:CK-SAC}
			\begin{algorithmic}[1]
				\STATE $\theta_{sa}$, $\theta_{sc}$, $theta_k$, $\theta_d$.\\
				\FOR {each iteration}
				\FOR{each environment step $i$}
				\STATE $a_k \sim \pi_{\theta_{sa}}(a_k|s_k)$
				\STATE $s_{k+1}\sim p(s_{k+1}|s_k, a_k)$
				\IF{$i\,\,\% p\,\,==0$}
				\STATE $\mathcal{D} \gets \mathcal{D} \cup \left\{ \left( s_{i-p:i},a_{i-p:i},r\left( s_{i-p:i},a_{i-p:i} \right) ,s_{i-p+1:i+1} \right) \right\} $
				\ENDIF
				\ENDFOR
				\FOR {each gradient step}
				\STATE $\theta _{sc}\gets \theta _{sc}-\beta _{sc}\triangledown _{\theta _{sc}}J_{V}$; \hfill{Update the critic}
				\STATE $\theta_K \gets \theta _{K} - \beta _{K}\triangledown _{\theta _{K}}L$	\hfill{Update the Koopman and AE}
				\STATE $\theta_{sa} \gets \theta _{sa} - \beta _{sa}\triangledown _{\theta _{sa}}J_{\pi}$ \hfill{Update the actor}
				\ENDFOR
				\ENDFOR
			\end{algorithmic}
		\end{algorithm}
	\end{appendices}
\end{document}